\documentclass[twocolumn]{autart}    

\usepackage[algo2e,norelsize,linesnumberedhidden,ruled]{algorithm2e}
\usepackage{marginnote}
\usepackage[english]{babel} 
\usepackage[utf8]{inputenc} 
\usepackage[cmex10]{amsmath} 
\usepackage{amssymb} 
\usepackage{float} 
\usepackage[pdftex]{graphicx} 
\usepackage{amsfonts} 
\usepackage{mathtools} 
\usepackage{xcolor} 
\usepackage{color} 
\usepackage{verbatim} 
\usepackage{graphicx} 
\usepackage{soul} 

{}
{}
{}
\newtheorem{theorem}{Theorem}{}
\newtheorem{remark}{Remark}{}
\newtheorem{lemma}{Lemma}{}
\newtheorem{structure}{Imposed Structure}{}

\begin{document}

\begin{frontmatter}

\title{Co-design for Security and Performance: LMI Tools\thanksref{footnoteinfo}} 

\thanks[footnoteinfo]{The authors are with the Departments of Mechanical and Systems Engineering at the University of Texas at Dallas.}

\author[Navid]{Navid Hashemi}\ead{nxh150030@utdallas.edu},    
\author[Justin]{Justin Ruths}\ead{jruths@utdallas.edu},               

\address[Navid]{The University of Texas at Dallas, Department of Mechanical Engineering}  
\address[Justin]{The University of Texas at Dallas, Departments of Mechanical and Systems Engineering}             


\begin{abstract}                    

We present a convex optimization to reduce the impact of sensor falsification attacks in linear time invariant systems controlled by observer-based feedback. We accomplish this by finding optimal observer and controller gain matrices that minimize the size of the reachable set of attack-induced states. To avoid trivial solutions, we integrate a covariance-based $\|H\|_2$ closed-loop performance constraint, for which we develop a novel linearization for this typically nonlinear, non-convex problem. We demonstrate the effectiveness of this linear matrix inequality framework through a numerical case study.
\end{abstract}

\end{frontmatter}

\section{Introduction}
A growing awareness of security concerns in automated physical processes has increased interest in our ability to quantify the impact of would-be attackers. Work along these lines imposes a detector to raise alerts when sensor measurements do not fall in line with model-based predictions, thus constraining what an attacker can do without being discovered \cite{Mo_3,milovsevic2018quantifying,RuthsACC2018_Comparison}. The tuning of such detectors is a balancing act between increasing the sensitivity to attacks while reducing the number of false alarms (alerts raised during normal operation) \cite{murguia2019model}. The next clear step in this direction of research is to then minimize this impact through careful control system design. Initially work used the distance (norm) that an attacker could drive the state as a proxy for impact \cite{murguia2019model,umsonst2018anomaly}, however, ultimately the reachable set provides important information about which components of a system are effected more than others and can inform whether the attacked state might reach dangerous regions of state space \cite{murguia2020security}. The quantification, or \textit{analysis}, studies have often used ellipsoidal bounds on the actual attack-induced reachable set achieved either through iterative methods \cite{Mo_3} or the satisfaction of linear matrix inequalities (LMIs) \cite{RuthsACC2018_Comparison,murguia2020security}. The latter extends gracefully to optimization to address the \textit{design} question, although doing so often requires re-linearizing the inequalities with respect to the new design variables.

In this paper, we leverage past work on quantification of ellipsoidal bounds on the attack-induced reachable set to design the observer and controller gain matrices to minimize the ellipsoidal bound (and thus the reachable set) when a linear time invariant system is controlled with estimate-based feedback. As has been pointed out (see, e.g., \cite{Carlos_Justin3,murguia2020security}), it is important to pair a security minimization with a constraint on closed-loop performance, otherwise a trivial solution exists to disconnect the feedback loop and thus cut off the effect of the attack on the system state. Here we specify an output covariance constrained (OCC) $\|H\|_2$ performance \cite{rotea1993generalized}, which takes into account the covariance of the noise, unlike other distribution-agnostic approaches to robust control \cite{de2002extended,geromel1999h}. Because we are able to know the distribution of the noises in our system, using them in the selection of optimal gains allows us to exploit this known structure for a tighter and more tailored result. The problem of covariance-based $\|H\|_2$ design is, however, in general a non-convex problem. To make it compatible with our LMI framework, along the way we develop a novel convexification of the OCC $\|H\|_2$ design problem, by dividing the problem into two parts: a convex optimization and a generalized algebraic Ricatti equation.

In \cite{Carlos_Justin3} we presented the notion of observer gain design to minimize the ellipsoidal bound on the estimation error. Here we complete the design problem by extending the framework to quantify and minimize the reachable set of attack-induced states, importantly integrating the design of the controller gain matrix as well. The more recent work \cite{murguia2020security} provides a general and expansive framework for using an LMI approach to solve the gain design problem for security. Our work here is distinguished by (a) using estimate feedback as opposed to a dynamic controller, (b) using a covariance-informed $\|H\|_2$ performance metric as opposed to a distributionally insensitive version, and (c) introducing a magnification factor to scale the shape matrices associated with the $\|H\|_2$ and reachable set decision variables. While the dynamic controller - characterized by a linear time invariant dynamical system with matrices $A^c$, $B^c$, $C^c$, and $D^c$ - used in \cite{murguia2020security} is a more general approach, the nonlinearities caused by the estimate-based feedback adds additional complications to the steps to linearize the constraints. In particular, if we define $A^c$, $B^c$, $C^c$, and $D^c$ to encode estimate-based feedback, the resulting inequalities in \cite{murguia2020security} are nonlinear and non-convex. The magnification factor we introduce here is effectively assumed to be unit value in \cite{murguia2020security}; in our numerical studies the value of this parameter tends to be large, indicating that including this factor greatly improves the quality of the optimization solution. A conference paper \cite{HashemiACC2020_LMI} presented the results for the iterative approach we take here, however, due to space the full detail of the proofs were not included. The fully convexified approach is not present in that paper.

\section{Background}
We consider a discrete-time linear time invariant (LTI) system of the form
\begin{align}\label{eq:dLTIsystem}
	x_{k+1} &= Fx_k + Gu_{k} + \nu_{k},\\
	y_k &= Cx_k + \eta_k,
\end{align}
in which the state $x_k\in\mathbb{R}^n$, $k\in\mathbb{N}$, evolves due to the state update provided by the state matrix $F\in\mathbb{R}^{n\times n}$, the control input $u_k\in\mathbb{R}^m$ shaped by the input matrix $G\in\mathbb{R}^{n\times m}$, and the i.i.d Gaussian system noise $\nu_k \sim \mathcal{N}(0,R_1)$, $R_1 \in \mathbb{V}^{n \times n}$ ($\mathbb{V}$ is the set of positive definite matrices). The output $y_k\in\mathbb{R}^p$ aggregates a linear combination of the states, given by the observation matrix $C\in\mathbb{R}^{p\times n}$, and the i.i.d Gaussian measurement noise $\eta_k \sim \mathcal{N}(0,R_2)$, $R_2 \in \mathbb{V}^{p \times p}$. For the simplicity of the exposition, we have considered Gaussian noises, however, the approach we present is applicable for general noise distributions. In addition we assume that $F$ is stable, the pair $(F,C)$ is detectable, $(F,G)$ is stabilizable and system and measurement noises are mutually independent.



In this work, we consider the scenario that the actual measurement $y_k$ can be corrupted by an attack, $\delta_k \in \mathbb{R}^p$. The attack is injected at some point between the measurement and reception of the output by the controller, 
\begin{equation}
	\bar{y}_k = y_k + \delta_{k} =  Cx_{k} + \eta_{k} + \delta_k.
\end{equation}
If the attacker has access to the measurements, then it is possible for the attack $\delta_k$ to cancel some or all of the original measurement $y_k$ - so an additive attack can achieve arbitrary control over the ``effective'' output of the system.

Because our system is stochastic, we require an estimator to produce a prediction of the system behavior
\begin{equation}
	\hat{x}_{k+1} = F\hat{x}_k + Gu_k + L(\bar{y}_k - C\hat{x}_k),
\end{equation}
where $\hat{x}_k \in \mathbb{R}^n$ is the estimated state and the observer gain $L$ is designed to force the estimate to track the system states.

We consider observer-based feedback controllers
\begin{equation}
    u_k=K\hat{x}_k,
\end{equation}
where $K \in \mathbb{R}^{m \times n}$ is the controller gain matrix. Next, we define the residual sequence
\begin{equation}
	r_k = \bar{y}_k - C\hat{x}_k,
\end{equation}
as the difference between what we actually receive ($\bar{y}_k$) and expect to receive ($C\hat{x}_k$), which evolves according to
\begin{equation}\label{eq:esterror}
\begin{aligned}
    &x_{k+1}=(F+GK)x_k-GKe_k+\nu_k\\
	&e_{k+1} = \big( F - LC \big) e_k - L\eta_k + \nu_k - L \delta_k, \\
	&r_k = Ce_k + \eta_k + \delta_k,
\end{aligned}
\end{equation}
where $e_k= x_k - \hat{x}_k$ is the estimation error. In the absence of attacks (i.e., $\delta_k = 0$), we can show that the steady-state distribution of $r_k$ is Gaussian with covariance,
\begin{equation}
\begin{aligned}\label{eq:Sigmacov}
	\Sigma &= \mathbf{E}[r_kr_k^T] = C\mathbf{E}[e_ke_k^T]C^T+\mathbf{E}[\eta_k\eta_k^T],\\
	&=C\mathbf{P}_eC^T+R_2,
\end{aligned}
\end{equation}
where the steady state covariance of the estimation error $\mathbf{P}_e= \lim_{k \rightarrow \infty}P_k= \lim_{k \rightarrow} \mathbf{E}[e_ke_k^T]$ is the solution of
\begin{equation}\label{eq:estcov}
\begin{aligned}
	&\mathbf{P}_e
	=(F-LC)\mathbf{P}_e(F-LC)^T +LR_2L^T+R_1.
\end{aligned}
\end{equation}
 In this work, we consider the chi-squared detector, although similar analysis can be done with other detector choices \cite{murguia2019model,umsonst2018anomaly}. The chi-squared detector constructs a quadratic distance measure $z_k$ to be sensitive to changes in the variance of the distribution as well as the expected value, 
\begin{equation}
	z_k = r_k^T\Sigma^{-1}r_k.
\end{equation}
The chi-squared detector generates alarms when the distance measure exceeds a threshold $\alpha \in \mathbb{R}_{>0}$
\begin{equation} \label{chisquared}
\left\{\begin{aligned}
	z_k \leq \alpha &\quad\longrightarrow\quad \text{no alarm}, \\
	z_k > \alpha &\quad\longrightarrow\quad \text{alarm: }k' = k,
\end{aligned}\right.
\end{equation}
such that alarm time(s) $k'$ are produced. The $\Sigma^{-1}$ factor in the definition of $z_k$ re-scales the distribution ($\mathbf{E}[z_k]=p$, $\mathbf{E}[z_kz_k^T]=2p$) so that the threshold $\alpha$ can be designed independent of the specific statistics (mean and covariance) of the noises $\nu_k$ and $\eta_k$; instead, it can be selected simply based on the number of sensors, $p$ \cite{murguia2019model}.

\subsection{Definition of Attack}
Detectors are designed to identify anomalies in system behavior. If an attacker aims to remain undetected, the choice of detector and its parameters limit what the attacker is able to accomplish. The type of attacks we consider here require strong knowledge of and access to system dynamics, statistics of the noises, current estimate ($\hat{x}_k$), and the detector configuration. The goal of this powerful stealthy attack is to construct the worst case scenario to aid the design of more robust systems.

\textit{Zero-alarm attacks} employ attack sequences that maintain the distance measure at or below the threshold of detection, i.e., $z_k \leq \alpha$. Hence, these attacks generate no alarms. To satisfy this condition we define the attack as
\begin{equation}\label{eq:deltabar}
	\delta_k =\phi_k-({y}_k-C\hat{x}_k)= - Ce_k-\eta_k+\phi_k,
\end{equation}
where $\phi_k\in\mathbb{R}^p$ is any vector such that $\phi_k^T\Sigma^{-1}\phi_k \leq \alpha$ (recall the attacker has access to the sensor, $y_k$, and knowledge of the estimator, $\hat{x}_k$). Based on this attack strategy,
\begin{align}
	z_k &= r_k^T\Sigma^{-1}r_k, \nonumber\\
    &= (Ce_k+\eta_k+\delta_k)^T\Sigma^{-1}(Ce_k+\eta_k+\delta_k), \nonumber\\
    &= \phi_k^T\Sigma^{-1}\phi_k \leq \alpha. \label{eq:zeroalarmdef}
\end{align}
Thus $z_k\leq\alpha$ and no alarms are raised.

\subsection{Reachable Set}
Under a stealthy zero-alarm attack \eqref{eq:deltabar}, the attacked system dynamics become
\begin{equation}
\begin{aligned}
    x_{k+1}&=Fx_k+GK\hat{x}_k+\nu_k,\\
    \hat{x}_{k+1}&=LCx_k+(F+GK-LC)\hat{x}_k-LCe_k+L\phi_k,\\
	e_{k+1} &=  Fe_k - L\phi_k + \nu_k.
\end{aligned}
\end{equation}
We stack these into a combined state $\xi_k=\left[x_k^T,\, \hat{x}_k^T,\, e_k^T\right]^T$ and combined input $\mu_k=\left[\nu_k^T,\, \phi_k^T\right]^T$,
\begin{equation}\label{eq:stackedattacked}
\xi_{k+1}=A\xi_{k}+B \mu_k,
\end{equation}
with
\begin{equation} \label{eq:AB}
{
     A=\begin{bmatrix}F & GK & 0\\ LC & F+GK-LC & -LC\\0 & 0 & F\end{bmatrix}, \   B=\begin{bmatrix}I & 0\\0 & L\\I& -L\end{bmatrix}.}
\end{equation}
\begin{remark}\label{rmk:redundantxi}
The choice of including $x_k$, $\hat{x}_k$, and $e_k$ seems redundant at this point since $e_k=x_k-\hat{x}_k$, however, this choice is crucial as we layer additional constraints into the design optimization.
\end{remark} 
Throughout the rest of the paper we will use a selection matrix $E_x=\left[I_{n},\, 0_{n\times n},\,  0_{n \times n}\right]$ to pull out quantities relevant to the state $x_k=E_x\xi_k$.

The reachable set of attack-induced states is then,
\begin{equation}
\mathcal{R} = \left\{  x_k=E_x\xi_k \ \left|\
    \begin{aligned}
        &\xi_{k+1}=A\xi_{k}+B \mu_k, \\
        &\xi_1={0},\ \phi_k^T\Sigma^{-1}\phi_k \leq \alpha,\\
        &\nu_k^TR_1^{-1}\nu_k \leq \bar{\nu},\ \forall k\in\mathbb{N}
    \end{aligned}
    \right. 
    \right\},
\end{equation}
where the ellipsoidal bound on the attack $\phi_k$ is imposed by the attacker's desire to remain stealthy \eqref{eq:zeroalarmdef}, and the ellipsoidal bound on the noise is created by truncating the Gaussian system noise to a desired probability, i.e., $\text{Pr}[\nu_k^TR_1^{-1}\nu_k \leq \bar{\nu}] = p_\nu$, where $p_\nu$ is some desired (typically high) probability. In principle, the noise has unbounded support, and hence the reachable set is unbounded. To ensure bounded reachable sets, we apply this truncation at the desired confidence level.

\section{LMI Approach to Design $K$ and $L$}
In the first section, we reframe an existing result more concisely, which identifies a minimal outer ellipsoidal bound on the set of states reachable by a stealthy (zero-alarm) attacker. We then move to consider minimizing this set further through the design of the feedback and estimator gains $K$ and $L$. As has been discussed in previous studies, a trivial solution exists to this design problem - to make either $GK=0$ or $L=0$. Doing so cuts the feedback loop and guarantees that corrupted measurements do not impact the system state. Simultaneously, this destroys the purpose - more specifically the performance - of the feedback loop. While many performance metrics could be used, in Section \ref{sec:h2} we impose a $\|H\|_2$ constraint to avoid these trivial solutions. Unlike prior work where the performance criteria ignored the distribution of the noise, this $\|H\|_2$ constraint is specific to the covariance of the noise, thereby allowing our design optimization to leverage this important knowledge. This output covariance constrained (OCC) $\|H\|_2$ constraint is non-convex; to our knowledge, this paper offers the first convexification of the OCC $\|H\|_2$ criteria into an LMI framework.


\subsection{Bounding Ellipsoid LMI (given $K$ and $L$)}
Before we move on to the \textit{synthesis} problem of designing the gain matrices, we first provide a solution to the \textit{analysis} problem of finding a tight outer ellipsoidal bound of the reachable set given $K$ and $L$, when the system is driven by the system noise and attack. A similar analysis result appears in \cite{RuthsACC2018_Comparison}, however, there the problem is split into two optimizations - one to find a bound on the estimation error reachable set, the result of which is used in the second optimization to bound the state reachable set. Here, in Lemma \ref{lem:analysis}, we solve these simultaneously through the stacked states $\xi_k$ and inputs $\mu_k$. The following lemma provides a bound on a Lyapunov-inspired function given an elliptically bounded input.

\begin{lemma}\cite{murguia2020security}\label{fatlemma}
Let $V_k$ be a positive definite function with $V_1 = 0$ and $\mu_{ik}^T W_i \mu_{ik} \leq 1$, $i=1, \dots, N$, where $W_i$ is positive definite. If there exists a constant $a \in (0,1)$ and $a_i \in (0,1)$  such that $\sum_{i=1}^{N} a_i \geq a$ and
\begin{equation}\label{eq:that}
V_{k+1} - aV_k -\sum_{i=1}^{N} (1-a_i) \mu_{ik}^T W_i\mu_{ik} \leq 0,
\end{equation}
then $V_k \leq \frac{N-a}{1-a}$.
\end{lemma}

When we select the positive definite $V_k$ to be a quadratic function of the state, the result above provides an outer ellipsoidal bound on the reachable states. We will use the notation $\mathcal{E}(\mathcal{Q})=\{x\ |\ x^T\mathcal{Q}^{-1}x\leq 1$\}.

\begin{lemma} \label{lem:analysis}
Given the stacked system matrices $A$ and $B$ in \eqref{eq:AB}, gain matrices $K$, $L$, detector threshold $\alpha$ with steady state residual covariance $\Sigma$, system noise truncation threshold $\bar{\nu}$ with covariance $R_1$, if there exists constants $a\in [0,1)$, the solution of 
\begin{equation} \label{eq:analysis}
\left\{ \begin{aligned}
\min_{a_1,a_2, \mathcal{Q}}  &\mathbf{tr}(E_x^T \mathcal{\mathcal{Q}} E_x)\\
\text{s.t.} &\ \ 0\leq a_1,a_2 < 1, \quad a_1+a_2 \geq a, \\
&\ \, \begin{bmatrix}a \mathcal{Q} &  \mathcal{Q} {A}^T &0\\ A \mathcal{Q} & \mathcal{Q} & B\\0 &  {B}^T & \frac{1-a}{2-a}W\end{bmatrix} \geq 0,
\end{aligned}\right.
\end{equation}
provides the shape matrix $\mathcal{Q}$ of the ellipsoidal bound on the reachable set of states, i.e., $\mathcal{R}\subseteq\mathcal{E}(E_x^T \mathcal{Q} E_x)$, where
\begin{equation}
    W=\begin{bmatrix}\frac{1-a_1}{\bar{\nu}}R_1^{-1} & 0 \\ 0 & (1-a_2)\Pi \end{bmatrix}, \ \ \Pi=\frac{\Sigma^{-1}}{\alpha}.
\end{equation}
\end{lemma}

\textit{Proof:} The stacked dynamics \eqref{eq:stackedattacked} is driven by two inputs which are both ellipsoidally bounded. Letting $W_1=R_1^{-1}$ and $W_2=\Sigma^{-1}$, \eqref{eq:that} becomes
\begin{equation}\label{eq:errorlemma2}
V_{k+1} - aV_k - \frac{1-b}{\alpha} \phi_k^T \Sigma^{-1} \phi_k -\frac{1-a_1}{\bar{\nu}}\nu_k^TR_1^{-1}\nu_k \leq 0.
\end{equation}
Substituting the choice $V_k=\xi_k^T \big(\frac{2-a}{1-a}\mathcal{P}\big) \xi_k \leq \frac{2-a}{1-a}$, $\mathcal{P}>0$, into this equation and expanding using the dynamics \eqref{eq:stackedattacked} results in the LMI,
\begin{equation}\label{eq:boundLMI2}
    \mathcal{H}=\begin{bmatrix}a\mathcal{P} &  {A}^T\mathcal{P}&0\\
    \mathcal{P} {A} & \mathcal{P} & \mathcal{P}  {B}\\
    0 &  {B}^T\mathcal{P} & \frac{1-a}{2-a}W
    \end{bmatrix}\geq 0,
\end{equation}
where $\mathcal{P}$ is the inverse of the shape matrix of the ellipsoidal bound for the $\xi$ reachable set ($\mathcal{P}^{-1}=\mathcal{Q}$), such that the first block $E_x^T\mathcal{Q}E_x$ is the shape matrix of the ellipsoidal bound of the reachable set of system states. To make this ellipsoidal bound tight (as small as possible), the cost is selected to minimize the trace of the shape matrix $E_x^T\mathcal{Q}E_x$. To use $\mathcal{Q}$ as the variable of the optimization instead of $\mathcal{P}$ we apply the transformation $T=\text{diag}\left[\mathcal{Q},\, \mathcal{Q},\, I_n \right]$, to \eqref{eq:boundLMI2}, i.e., $T^T \mathcal{H} T$, which results in the LMI in \eqref{eq:analysis}.
\hspace*{\fill}~$\blacksquare$

\begin{remark} \label{rmk:traceobjective}
For any convex shape (i.e., reachable set) there are an infinite number of \textit{tight} outer ellipsoidal bounds. These different ellipsoids can be visualized as being tangent to the reachable set at different points. The minimum trace objective minimizes the sum of the squared principal axes, which tends to avoid solutions with, for example, low volume but one large principal axis. 
\end{remark}

\begin{remark}
Note that the parameter $a$ is not a decision variable of the optimization in \eqref{eq:analysis}. It appears nonlinearly (multiplying $\mathcal{Q}$). Since $a$ belongs to a compact interval, the conventional choice is to solve \eqref{eq:analysis} across a grid search in $a$ and select the minimal, feasible solution.
\end{remark}

\subsection{Output Covariance Constrained $\|H\|_2$ Constraint} \label{sec:h2}
The introduction of this section and past related work has identified that trivial solutions exist for the synthesis problem unless a performance criteria is imposed in the optimization \cite{Carlos_Justin3,murguia2020security}. One of the distinguishing features of this work is that we consider an output covariance constrained $\|H\|_2$ constraint, which involves the covariances of the system and sensor noises. The challenge, tackled in the next subsection, is to convexify and linearlize this inherently nonlinear constraint. Most optimizations in the literature either use a distributionally robust constraint that is already convex \cite{murguia2020security,de2002extended} or solve the OCC $\|H\|_2$ using iterative algorithms \cite{zhu1997convergent,xu2006monitoring}.  
To specify the performance, Robust Control, in general, studies the gain observed in the signal $h_k=H_1x_k+H_2\eta_k+H_3\nu_k$. Here, for the system \textit{without attack}, we consider the system driven by system and measurement noise and enforce an $\|H\|_2$ constraint between the output $h_k=y_k$ and excitation $\omega_k= \left[\nu_k^T,\, \eta_k^T \right]^T$, making $H_1=C$, $H_2=I_{p\times p}$, and $H_3=0_{p \times n}$.

When there is no attack the system evolves according to
 \begin{align}
    x_{k+1}&=Fx_k+GK \hat{x}_k+\nu_k,\label{eq:statenoattack}\\
	\hat{x}_{k+1}&=LCx_k+(F+GK-LC)\hat{x}_k+L\eta_k, \label{eq:estimationnoattack}\\
	y_k &= Cx_k + \eta_k, \label{eq:outputnoattack}
\end{align}
which can be combined using the stacked state $\zeta_k=E_{x\hat{x}}\xi_k=\left[x_k^T,\, \hat{x}_k^T\right]^T$,
\begin{equation}\label{eq:stackednoattacked}
\zeta_{k+1}=\hat{A}\zeta_k+\hat{B}\omega_k,
\end{equation}
with $E_{x\hat{x}}=\left[I_{2n},\, 0_{2n \times n}\right]$ making $\hat{A}=E_{x\hat{x}}AE_{x\hat{x}}^T$ and $\hat{B}=E_{x\hat{x}}B E_{x\hat{x}}^T$.

\begin{remark}\label{rmk:paralleldynamics}
It is here that we can start to appreciate the value of the seemingly redundant definition of $\xi_k$ (see Remark \ref{rmk:redundantxi}). By doing so, the state matrix without attack $\hat{A}$ can be expressed as a sub-block of the state matrix under attack $A$. Establishing this parallel structure is key towards being able to integrate the $\|H\|_2$ constraint (without attack) with the reachable set calculation (under attack).
\end{remark}

The OCC $||H||_2$ criteria specifies the gain from the noise to the output should be less than a desired value $\bar{\gamma}$,
\begin{equation} \label{eq:originalineq}
\lim_{N \to \infty}\sqrt{\frac{\frac{1}{N}\sum_{k=1}^N y_k^Ty_k}{\frac{1}{N}\sum_{k=1}^N\omega_k^T\omega_k}} = \sqrt{\frac{\mathbf{E}[y_k^Ty_k]}{\mathbf{E}[\omega_k^T\omega_k]}} \leq \bar{\gamma}.
\end{equation}
\begin{lemma}
Given the dynamics in \eqref{eq:stackednoattacked}, the OCC $\|H\|_2$ constraint in \eqref{eq:originalineq} is satisfied if the steady state covariance 
\begin{equation}
    \mathbf{P}=\begin{bmatrix}\mathbf{P}_{x}&\mathbf{P}_{x\hat{x}}\\\mathbf{P}_{x\hat{x}}^T&\mathbf{P}_{\hat{x}}\end{bmatrix} = \lim_{k\to\infty}\mathbf{P}_{k} = \lim_{k\to\infty}\mathbf{E}[\zeta_k\zeta_k^T],
\end{equation}
satisfies the Lyapunov equation
\begin{equation}\label{eq:xcovariance}
    \mathbf{P}=\hat{A} \mathbf{P} \hat{A}^T+\hat{R},\qquad \mathbf{P}\geq 0,
\end{equation}
and the following {convex} inequality holds,
\begin{equation} \label{eq:Hinfty}
\begin{aligned}
\mathcal{C}_h=\mathbf{tr} \big( \hat{E}_x^TC^TC\hat{E}_x &\mathbf{P} \big) + \mathbf{tr} (R_2) \\&- \bar{\gamma}^2 \big(\mathbf{tr}(R_1)+\mathbf{tr}(R_2) \big) \leq 0,
\end{aligned}
\end{equation}
where $\hat{E}_x=\left[I_n,\, 0_{n \times n}\right]$.
\end{lemma}

\textit{Proof:}
From \eqref{eq:outputnoattack} and the definition of $\omega_k$, we can calculate the quadratic terms in \eqref{eq:originalineq},
\begin{align}
y_k^T y_k &=  x_k^TC^T  C x_k +  2 x_k^TC^T \eta_k + \eta_k^T \eta_k,\\
\omega_k^T\omega_k&=\nu_k^T \nu_k+\eta_k^T\eta_k.
\end{align}
Taking the expectation ($x_k$ and $\eta_k$ are independent), 
\begin{equation}\label{eq:yexpectation}
\begin{aligned}
\mathbf{E}[y_k^T y_k] &=  \mathbf{E}[x_k^T C^T  C x_k] +\mathbf{E}[\eta_k^T \eta_k]\\
&=  \mathbf{tr}\big( C^TC \ \mathbf{E}[x_kx_k^T] \big)  + \mathbf{tr}(R_2), 
\end{aligned}
\end{equation}
 and similarly,
\begin{equation}\label{eq:omegaexpectation}
\mathbf{E}[\omega_k^T\omega_k]=\mathbf{E}[\nu_k^T \nu_k] +\mathbf{E}[\eta_k^T\eta_k]=\mathbf{tr}(R_1)+\mathbf{tr}(R_2).
\end{equation}
The unknown quantity is then the covariance of the state, $\mathbf{E}[x_kx_k^T]$, which is the first block of the stacked state $\zeta_k$ covariance $\mathbf{P}_k=\mathbf{E}[\zeta_k\zeta_k^T]$. This covariance follows the update, evaluating $\mathbf{E}[\zeta_{k+1}\zeta_{k+1}^T]$ with \eqref{eq:stackednoattacked},
\begin{equation}
\mathbf{P}_{k+1}=\hat{A} \mathbf{P}_{k} \hat{A}^T+\hat{R},\qquad \hat{R}=\begin{bmatrix}R_1 & 0\\0 & LR_2L^T\end{bmatrix}.
\end{equation}
Because the matrix $\hat{A}$ is stable the covariance converges to a steady value $\lim_{k\to\infty}\mathbf{P}_{k} = \mathbf{P}$ which satisfies the Lyapunov equation \eqref{eq:xcovariance}.
Combining this $\mathbf{P}$ with \eqref{eq:yexpectation}-\eqref{eq:omegaexpectation} the $\|H\|_2$ constraint becomes
\begin{equation}\label{eq:nonconvexgamma}
   \gamma=\sqrt{\frac{ \mathbf{tr} \big( C \mathbf{P}_x C^T \big) + \mathbf{tr}(R_2) }{\mathbf{tr}(R_2)+\mathbf{tr}(R_1)}}\leq \bar{\gamma},
\end{equation}
where $\gamma$ is the actual performance and $\bar{\gamma}$ is the worst allowable performance. Rearranging this leads to the condition \eqref{eq:Hinfty}.
\hspace*{\fill}~$\blacksquare$

In order to use this $\|H\|_2$ constraint in a convex optimization we need to linearize the Lyapunov equation constraint. We state this result as part of a complete convex optimization problem to design the gains $K$ and $L$ to achieve the optimal (smallest) $\|H\|_2$ gain.

\begin{theorem}\label{covbaseH}
Given the dynamics in \eqref{eq:stackednoattacked}, the smallest output covariance constrained $\|H\|_2$ gain defined by \eqref{eq:originalineq} is  
\begin{equation}\label{eq:optimgamma}
\gamma^*=\sqrt{\frac{ \mathbf{tr} \big( C \mathbf{P}_x^* C^T \big) + \mathbf{tr}(R_2) }{\mathbf{tr}(R_2)+\mathbf{tr}(R_1)}},
\end{equation}
where $\mathbf{P}_x^*$ is the solution of
\begin{equation}\label{eq:gammaopt}
\left\{\begin{aligned}
\min_{\mathbf{P}_x,\mathbf{Q}_1,X,Y,Z} & \mathbf{tr}(C\mathbf{P}_xC^T)\\
\text{s.t.}\quad & \mathcal{C}_L \geq 0,
\end{aligned}\right.
\end{equation}
with
\begin{equation} \label{eq:linearizedH2Lyapunov}
   \mathcal{C}_L = \resizebox{0.75\hsize}{!}{$\begin{bmatrix}\mathbf{Q}_1 & I & \mathbf{Q}_1F+XC & Z & \mathbf{Q}_1 R_1 & XR_2\\
    * & \mathbf{P}_x & F & F\mathbf{P}_x+GY & R_1 & 0 \\
    * & * & \mathbf{Q}_1 & I & 0 &0 \\
    * & * & * & \mathbf{P}_x & 0 & 0 \\
    * & * & * & * & R_1 & 0\\
    * & * & * & * & * & R_2\end{bmatrix}$}.
\end{equation}
There exists at most $2n \choose n$ distinct real-valued, control gains $L=\mathbf{Q}_{12}^{-1}X$, $K=Y\mathbf{P}_{x\hat{x}}^{-T}$ satisfying $\gamma=\gamma^*$, where $\mathbf{P}_{x\hat{x}}=(I-\mathbf{P}_x \mathbf{Q}_1)\mathbf{Q}_{12}^{-T}$ and $\mathbf{Q}_{12}$ is the solution of following generalized algebraic Ricatti equation,
\begin{equation}\label{eq:genricatti}
\mathbf{Q}_{12}\Gamma_1\mathbf{Q}_{12}+\mathbf{Q}_{12}\Gamma_2+\Gamma_3\mathbf{Q}_{12}+\Gamma_4=0,
\end{equation}
with known matrices
\begin{align}
&\Gamma_1=GY(I-\mathbf{Q}_1\mathbf{P}_x)^{-1},\\
&\Gamma_2=F,\nonumber\\
&\Gamma_3=(\mathbf{Q}_1GY+XC\mathbf{P}_x+\mathbf{Q}_1F\mathbf{P}_x-Z)(I-\mathbf{Q}_1\mathbf{P}_x)^{-1},\nonumber \\
&\Gamma_4=-XC.\nonumber
\end{align}
\end{theorem}
\textit{Proof:}
The formula for the optimal gain $\gamma^*$, \eqref{eq:optimgamma}, comes naturally from the $\|H\|_2$ bound derived in \eqref{eq:nonconvexgamma}. Since all other terms are constant, minimizing $\mathbf{tr}(C\mathbf{P}_xC^T)$ is equivalent to minimizing the gain. This covariance is constrained by the Lyapunov equation in \eqref{eq:xcovariance}. Here, which is a standard technique for incorporating Lyapunov equations into convex optimizations, we replace this equality constraint with the very similar inequality, 
\begin{equation}\label{eq:replacement}
\mathbf{P}-\hat{A} \mathbf{P} \hat{A}^T-\hat{R} \geq 0,\qquad \mathbf{P}\geq0.
\end{equation}
We can now combine these two inequality constraints into one using the Schur complement \cite{BEFB:94},
\begin{equation}
   \mathcal{C}= \begin{bmatrix}\mathbf{P}-\hat{R} & \hat{A}\mathbf{P} \\ \mathbf{P}\hat{A}^T &\mathbf{P}\end{bmatrix} \geq 0.
\end{equation}
This relaxation is justified because the objective function $\mathbf{tr}(C\mathbf{P}_xC^T)$ minimizes the decision variable $\mathbf{P}$ and drives the optimization to the bound of the inequality, which would yield equality - hence driving the relaxed form \eqref{eq:replacement} to the equality \eqref{eq:xcovariance} (see Appendix \ref{apdx:lyapunov_relax}).

We use the following transformation to linearize $\mathcal{C}$,
\begin{equation}
    \mathcal{C}_L=\begin{bmatrix}T_1&\\&T_1\end{bmatrix}^T\mathcal{C}\begin{bmatrix}T_1&\\&T_1\end{bmatrix}=\begin{bmatrix}\mathbf{P}_L-R_L & \hat{A}_L \\\hat{A}_L^T & \mathbf{P}_L\end{bmatrix},
\end{equation}
with
\begin{equation} \label{eq:PinvQ}
    T_1=\begin{bmatrix}\mathbf{Q}_1 & I \\\mathbf{Q}_{12}^T & 0\end{bmatrix},\qquad 
    \mathbf{P}^{-1}=\mathbf{Q}=\begin{bmatrix}\mathbf{Q}_1&\mathbf{Q}_{12}\\\mathbf{Q}_{12}^T&\mathbf{Q}_2\end{bmatrix},
\end{equation}
and
\begin{align}
&\mathbf{P}_L=T_1^T \mathbf{P} T_1=\begin{bmatrix}\mathbf{Q}_1 & I \\ I & \mathbf{P}_x\end{bmatrix}, \\
&R_L=T_1^TRT_1=\begin{bmatrix}\mathbf{Q}_1R_1\mathbf{Q}_1\ +\ \mathbf{Q}_{12}LR_2L^T\mathbf{Q}_{12}^T & \mathbf{Q}_1R_1\\ R_1\mathbf{Q}_1 & R_1\end{bmatrix},\nonumber\\
&\hat{A}_L=T_1^T\hat{A} \mathbf{P} T_1=\begin{bmatrix}\mathbf{Q}_1F+XC&Z\\F&F\mathbf{P}_x+GY\end{bmatrix},\nonumber
\end{align}
where $X$, $Y$, and $Z$ are defined as 
\begin{align}
    X&=\mathbf{Q}_{12}L, \label{eq:X}\\
    Y&=K\mathbf{P}_{x\hat{x}}^T, \label{eq:Y}\\
    Z&=\mathbf{Q}_1F\mathbf{P}_x+XC\mathbf{P}_x+\mathbf{Q}_1GY+\mathbf{Q}_{12}F\mathbf{P}_{x\hat{x}}^T \label{eq:Z}\\
    &\qquad\qquad+\mathbf{Q}_{12}GY-XC\mathbf{P}_{x\hat{x}}^T. \nonumber
\end{align}
The term $\mathbf{P}_L-R_L$ can be linearized by applying a Schur complement to recover $\mathcal{C}_L$ in \eqref{eq:linearizedH2Lyapunov}. This transformation changes the set of decision variables from ($\mathbf{P}_x$, $\mathbf{P}_{\hat{x}}$, $\mathbf{P}_{x\hat{x}}$, $L$, $K$) to ($\mathbf{P}_x$, $\mathbf{Q}_1$, $X$, $Z$, $Y$). The solution in these new decision variables is then used to calculate $\mathbf{P}_{x\hat{x}}$ and $\mathbf{Q}_{12}$ using \eqref{eq:Z} and the identity 
\begin{equation}\label{eq:UVanswerset}
\mathbf{P}_x\mathbf{Q}_1+\mathbf{P}_{x\hat{x}}\mathbf{Q}_{12}^T=I,
\end{equation}
which comes from the first block of the definition $\mathbf{P}\mathbf{Q}=I$. The definition of $Z$ in \eqref{eq:Z} and \eqref{eq:UVanswerset} combine to form the general algebraic Ricatti equation  \eqref{eq:genricatti}, which, in general, has $2n \choose n$ different answers. Finally, the gain matrices can be found by, $L=\mathbf{Q}_{12}^{-1}X$ and $K=Y\mathbf{P}_{x\hat{x}}^{-T}$. Note that the solutions to the original Lyapunov equality \eqref{eq:xcovariance} are a subset of the solutions of the relaxed inequality \eqref{eq:replacement}; so Theorem \ref{covbaseH} characterizes all OCC $\|H\|_2$ optimal solutions.
\hspace*{\fill}~$\blacksquare$

At the other end of the performance spectrum, security increases (size of the reachable set decreases) as either $L$ or $GK$ approach zero. This represents worst-case performance (without closed loop control) and the corresponding value of $\gamma=\gamma_0$ can be found using Lemma \ref{lem:openloopgamma}. 
\begin{lemma} \label{lem:openloopgamma}
Given the system dynamics,in the absence of attack, the open loop (i.e., $L=0$ or $GK=0$) OCC $\|H\|_2$ gain $\gamma_0$ is given by
\begin{equation*}
\gamma_0=\sqrt{\frac{ \mathrm{tr} \big(C \mathbf{P}_x  C^T \big) + \mathrm{tr}(R_2) }{\mathrm{tr}(R_2)+\mathrm{tr}(R_1)}},
\end{equation*}
where the steady state covariance of state $\mathbf{P}_x$ is the solution of
\begin{equation*} \label{eq:openlooplyapunov}
F\mathbf{P}_xF^T-\mathbf{P}_x+R_1=0.
\end{equation*}
\end{lemma}
\textit{Proof.}
For both $L=0$ or $GK=0$, the state dynamics \eqref{eq:statenoattack} are in open loop and thus the evolution of the system is the same. When $L=0$, the state estimate $\hat{x}_k$ converges to zero because the system is open loop stable and the open loop state dynamic becomes
\begin{equation}\label{eq:openloopstate}
    x_{k+1}=Fx_k+\nu_k.
\end{equation} 
Similarly, when $GK=0$, equation \eqref{eq:statenoattack} becomes \eqref{eq:openloopstate} directly. With this state equation, the steady state covariance of the state, $\mathbf{P}_x$, is given by \eqref{eq:openlooplyapunov} and consequently the desired performance  $\gamma_0$ should be the same in both cases.
\hspace*{\fill}~$\blacksquare$

Therefore, now entering into the design process, for all choices of $\bar{\gamma} \in [\gamma^*,\ \infty]$, the solution for performance $\gamma<\bar{\gamma}$ always lies within the trade-off interval, 
\begin{equation}\label{eq:tradeoffinterval}
\gamma^* \leq \gamma \leq \gamma_0.
\end{equation}


\subsection{Bounding ellipsoid LMI (designing $K$ and $L$)}
The goal of this paper is to construct an optimization to design $K$ and $L$ such that the impact of an attacker on the reachable states is minimized. However, when $K$ and $L$ are considered variables of the Lemma \ref{lem:analysis} optimization, \eqref{eq:boundLMI2}, and therefore, \eqref{eq:analysis} contains nonlinear terms. In the sections that follow, we impose some structure on the solution so that we can linearize the overall design problem. Each choice will be motivated individually, but it is also the combined effect of the these structures taken together that yield the final \textit{linear} matrix inequality. The first three (out of four) represent a choice of how to select an outer ellipsoidal bound. While they do impose some structure on the solution, they are best seen as one choice out of infinitely many equally good options, so they are not very limiting.

\begin{structure}
There are an infinite number of tight outer ellipsoidal bounds of the stacked state $\xi$. Of these we select one that satisfies the following structure for the inverse of the shape matrix,
\begin{equation}
    \mathcal{P}=\begin{bmatrix}\mathcal{P}_{1}&\mathcal{P}_{12}&\\\mathcal{P}_{12}^T& \mathcal{P}_{2} & \\&& \mathcal{P}_{3}\end{bmatrix},
\end{equation}
which assumes the independence of the ellipsoidal bound on the estimation error $e_k$ from the ellipsoidal bound on the combined state $x_k$ and estimate $\hat{x}_k$. This is inspired by a similar assumption made in \cite{murguia2020security}. This selection enables us to utilize the parallel dynamics with and without attack (see Remarks \ref{rmk:redundantxi} and \ref{rmk:paralleldynamics}) and linearize the original LMI with respect to $K$ and $L$.
\end{structure}

This selection also permits inverting each block separately, such that $\mathcal{P}_3^{-1}=\mathcal{Q}_e$ and
\begin{equation} \label{eq:shapematrix12}
    \begin{bmatrix}\mathcal{P}_{1}&\mathcal{P}_{12}\\\mathcal{P}_{12}^T& \mathcal{P}_{2}\end{bmatrix}^{-1}=
   \begin{bmatrix}\mathcal{Q}_x&\mathcal{Q}_{x\hat{x}}\\\mathcal{Q}_{x\hat{x}}^T&\mathcal{Q}_{\hat{x}}\end{bmatrix}.
\end{equation}
Consider the linearizing change of coordinates used in \cite{geromel1999h,murguia2020security},
\begin{equation}
T_2=\begin{bmatrix}T_3&&\\&T_3&\\&&I_n\end{bmatrix},\qquad
T_3= \begin{bmatrix}\mathcal{Q}_x & I & 0\\ \mathcal{Q}_{x\hat{x}}^T & 0 & 0 \\ 0 & 0 & I\end{bmatrix}.
\end{equation}
Although \eqref{eq:boundLMI2} is not entirely linearized with this transformation, due to the presence of term $\Sigma$ which depends on $L$, we will introduce an iterative approach later to avoid this nonlinearity. The LMI $\mathcal{H}$, \eqref{eq:boundLMI2}, becomes
\begin{equation}
    \begin{aligned}
    &\mathcal{H}_L=T_2^T\mathcal{H}T_2=\begin{bmatrix}a\mathcal{P}_L & A_L^T & 0\\A_L & \mathcal{P}_L & B_L\\ 0 & B_L^T & \frac{1-a}{2-a}W\end{bmatrix},
    \end{aligned}
\end{equation}
where
\begin{align}
    &\mathcal{P}_L=T_3^T\mathcal{P} T_3=\begin{bmatrix}\mathcal{Q}_x & I & 0\\I & \mathcal{P}_{1} & 0 \\ 0 & 0 & \mathcal{P}_{3}\end{bmatrix},\\
    &B_L=T_3^T\mathcal{P} B=\begin{bmatrix}I & 0 \\\mathcal{P}_{1} & Y_1\\\mathcal{P}_{3} & -\mathcal{P}_{3}L\end{bmatrix}, \nonumber\\
    &A_L=T_3^T\mathcal{P} AT_3=\begin{bmatrix}F\mathcal{Q}_x+GX_1 & F & 0\\Z_1 & \mathcal{P}_{1}F+Y_1C & -Y_1C\\0 & 0 & \mathcal{P}_{3}F\end{bmatrix}, \nonumber\\
    &Y_1=\mathcal{P}_{12}L, \quad X_1=K\mathcal{Q}_{x\hat{x}}^T,\nonumber\\
    &Z_1=\mathcal{P}_{1}F\mathcal{Q}_x+\mathcal{P}_{12}LC\mathcal{Q}_x+\mathcal{P}_{1}GK\mathcal{Q}_{x\hat{x}}^T+\mathcal{P}_{12}F\mathcal{Q}_{x\hat{x}}^T\nonumber\\
    &\qquad+\mathcal{P}_{12}GK\mathcal
    {Q}_{x\hat{x}}^T-\mathcal{P}_{12}LC\mathcal{Q}_{x\hat{x}}^T.\nonumber
\end{align}
One of the useful features of this transformation is that $\mathcal{Q}_x=E_x^T \mathcal{Q} E_x$, the quantify used in the objective function of Lemma \ref{lem:analysis}, appears as a variable of the LMI. This section provides the linearization necessary to separate the gains $K$ and $L$ as variables in Lemma \ref{lem:analysis} (and could then be used as the starting point if a different performance criteria was used, as opposed to the $\|H\|_2$ constraint considered in this paper). 

\begin{figure}[t]
    \centering
    \includegraphics[width=0.5\linewidth]{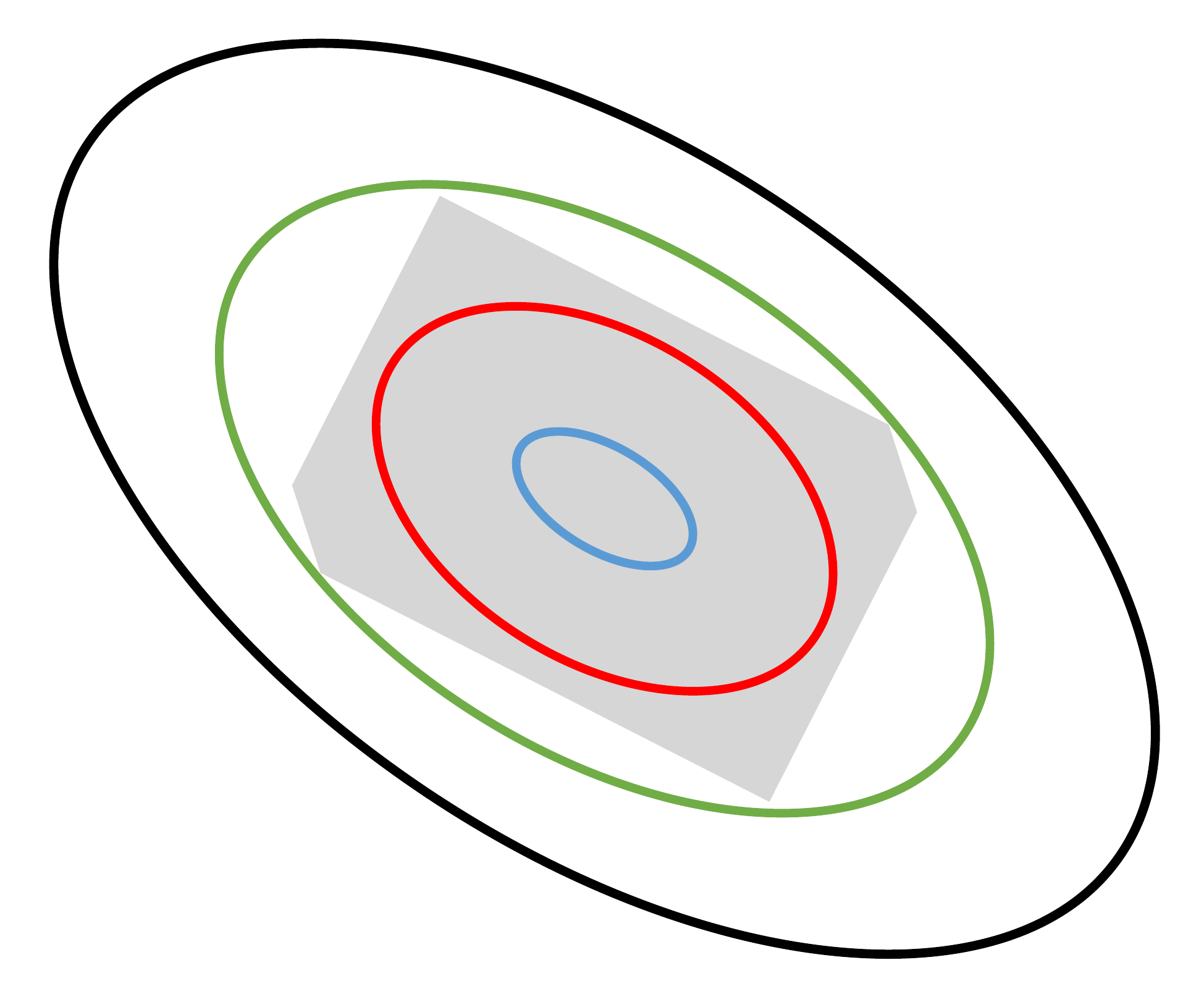}
    \caption{The reachable set (gray) is approximated by ellipsoids with shape matrix $\sigma\mathbf{P}$. Without relaxation effects, the role of $\sigma$ is to identify the tight approximation (green), where $\sigma=\sigma^*$. When $\sigma<\sigma^*$ the optimization will be infeasible or doesn't converge, because the red ellipsoid cannot contain the reachable set. When $\sigma>\sigma^*$ the black ellipsoid loosely contains the reachable set. Due to our techniques to linearize the optimization, the reachable set approximation will have some extra conservatism and we do not expect the optimal ellipsoid to be exactly tangent.}
    \label{fig:sigmashape}
\end{figure}

\subsection{Combining Performance and Security}
In this work, we design the controller and estimator gains to minimize the impact of attacks on the system state, which is measured by an outer ellipsoidal bound on the reachable states when the system is driven by the attack and system noise. As Remark \ref{rmk:traceobjective} states, there are an infinite number of potential outer bounding - and tight - ellipsoids. In order to combine the LMI constraints from the reachable set and $\|H\|_2$ calculations, we make a specific choice about the outer ellipsoidal bound we select.

\begin{structure} \label{structure:xxhat}
We select the shape matrix of the ellipsoidal bound of the states $x_k$ and estimate $\hat{x}_k$ under attack $E_{x\hat{x}}^T \mathcal{Q} E_{x\hat{x}}$ - see \eqref{eq:shapematrix12} - to have the same orientation as the covariance of the states and estimate without attack ($\zeta_k$), 
\begin{equation}\label{eq:wonderful}
    \sigma\mathbf{P}=E_{x\hat{x}}^T \mathcal{Q} E_{x\hat{x}},
\end{equation}
where $\sigma$ is a scaling factor that becomes a new variable of the method and is a function of gains $(L,\ K)$.  Since $\mathbf{Q} = \mathbf{P}^{-1}$, this sets up a common set of variables to link the $\|H\|_2$ (left) and ellipsoidal bound (right) constraints,
\begin{equation}  \label{eq:linearfactor}
\begin{aligned}
   \sigma \begin{bmatrix}\mathbf{P}_{x}&\mathbf{P}_{x\hat{x}}\\\mathbf{P}_{x\hat{x}}^T&\mathbf{P}_{\hat{x}}\end{bmatrix} &= \begin{bmatrix}\mathcal{Q}_x&\mathcal{Q}_{x\hat{x}}\\\mathcal{Q}_{x\hat{x}}^T&\mathcal{Q}_{\hat{x}}\end{bmatrix},\\
   \underbrace{\frac{1}{\sigma}\begin{bmatrix}\mathbf{Q}_1&\mathbf{Q}_{12}\\\mathbf{Q}_{12}^T&\mathbf{Q}_2\end{bmatrix}}_{\|H\|_2} &= \underbrace{\begin{bmatrix}\mathcal{P}_{1}&\mathcal{P}_{12}\\\mathcal{P}_{12}^T& \mathcal{P}_{2}\end{bmatrix}.}_{\mathcal{E}(\mathcal{Q})}
\end{aligned}\ 
\end{equation}
The structure above allows us to replace variables in the ellipsoidal bound optimization $\mathcal{Q}$ and $\mathcal{P}$ with quantities from the performance criteria, $\mathbf{P}$ and $\mathbf{Q}$, respectively.
\end{structure}

Based on \eqref{eq:linearfactor} we can link the variables of the bounding ellipsoid LMI with the $\| H\|_2$ constraint,
\begin{align}
X&=\sigma Y_1=\mathbf{Q}_{12}L, \quad Y=\frac{X_1}{\sigma}=K\mathbf{P}_{x\hat{x}}^T,\\
Z&=Z_1=\mathbf{Q}_1F\mathbf{P}_x+XC\mathbf{P}_x+\mathbf{Q}_1GY+\mathbf{Q}_{12}F\mathbf{P}_{x\hat{x}}^T \nonumber\\
&\qquad\qquad+\mathbf{Q}_{12}GY-XC\mathbf{P}_{x\hat{x}}^T. \nonumber
\end{align}
Now we can rewrite $A_L$, $B_L$ , $\mathcal{P}_L$ based on $\mathbf{P}_x$, $\mathbf{Q}_1$, $\mathcal{P}_3$, $X$, $Y$, $Z$,
\begin{equation}\label{eq:newboundmatrices}
\begin{aligned}
&A_L=\begin{bmatrix}\sigma (F\mathbf{P}_x+GY) & F & 0\\ Z & \frac{1}{\sigma}(\mathbf{Q}_1F+XC) & -\frac{1}{\sigma}XC\\ 0 & 0 & \mathcal{P}_3F\end{bmatrix},\\
&B_L=\begin{bmatrix}I&0\\\frac{1}{\sigma}\mathbf{Q}_1&\frac{1}{\sigma}X\\\mathcal{P}_3& -\mathcal{P}_3L\end{bmatrix}, \quad \mathcal{P}_L=\begin{bmatrix}\sigma \mathbf{P}_x &I&0\\I&\frac{1}{\sigma}\mathbf{Q}_1&0\\0&0&\mathcal{P}_3\end{bmatrix}.
\end{aligned}
\end{equation}
Thus the choice in \eqref{eq:wonderful} has facilitated integrating these optimizations.

\begin{algorithm2e}[t]
\DontPrintSemicolon
\caption{$K,L =$ Theorem \ref{semifinalconvex}$(F,G,C,R_1,R_2)$} \label{thealgorithm}
$L \leftarrow$ \text{such that $\mathcal{C}_L\geq 0$ is feasible}\;
$\sigma \leftarrow \infty$\;
\While{true} {
	    \lIf{\eqref{eq:semifinalconvex} is infeasible }{$\mathbf{break}$}
        $\mathbf{P}_x, \mathbf{Q}_1, X, Y, Z \leftarrow$ \eqref{eq:semifinalconvex} \;
        $\mathcal{K}, \mathcal{L} \leftarrow$ \text{real solutions of Ricatti eqn \eqref{eq:genricatti}}\;
        $(\tilde{K},\tilde{L}) \leftarrow$ select pair $(K,L)\in(\mathcal{K},\mathcal{L})$ by smallest\; \hspace{1.5cm}\textit{Lemma} \ref{lem:analysis} objective value\;
        \lIf{$\|L-\tilde{L}\|\leq \epsilon$}{$\sigma \leftarrow \sigma-\varepsilon$}
        \lElse{$K,L \leftarrow \tilde{K},\tilde{L}$}
}
\end{algorithm2e}

\begin{theorem}\label{semifinalconvex}
Consider a LTI system \eqref{eq:dLTIsystem} with maximum allowable output covariance constrained $\|H\|_2$ gain $\bar{\gamma}$ \eqref{eq:originalineq}, chi-squared detector threshold $\alpha$ \eqref{chisquared} and zero-alarm stealthy attacker \eqref{eq:zeroalarmdef}. Algorithm \ref{thealgorithm} returns (approximately) optimal controller $K^*$ and observer $L^*$ gains to minimize the reachable set of states possible by the attacker, while maintaining an OCC $\|H\|_2$ gain no bigger than $\bar{\gamma}$. Algorithm \ref{thealgorithm} uses the Ricatti equation \eqref{eq:genricatti} to update $L$ based on the solution of the combined convex optimization problem which has a solution if for some $a \in [0,1)$,
\begin{equation}\label{eq:semifinalconvex}
\begin{aligned}
&\left\{ \begin{aligned}
&\min_{{a_1,a_2,\mathbf{P}_x,\mathbf{Q}_1,}\atop{X,Y,Z,\mathcal{P}_3}}  \mathbf{tr}(\sigma \mathbf{P}_x)\\
&\qquad\text{s.t.}\quad 0\leq a_1,a_2 < 1, \quad a_1+a_2 \geq a,\\
&\qquad\qquad\mathcal{H}_L\geq0,\\
&\qquad\qquad\mathcal{C}_{h} \leq 0,\ \mathcal{C}_L \geq 0.
\end{aligned}\right.
\end{aligned}
\end{equation}
\end{theorem}

\textit{Proof:}
In the past sections we have linearized the LMIs associated with the ellipsoidal outer bound on the reachable set ($\mathcal{H}_L$) and with the $\|H\|_2$ constraint ($\mathcal{C}_h$ and $\mathcal{C}_L$) and finally made a structural connection between these two optimizations \eqref{eq:linearfactor} to use a common set of decision variables. Because the controller gain $K$ appears within the decision variables of the optimization, we implicitly optimize $K$.

There are two remaining challenges to be addressed by this algorithm. First, $L$ appears, as $K$ does, implicitly in the decision variables, but also explicitly in the nonlinear term $\mathcal{P}_3L$ in matrix $B_L$ and in the dependence of covariance $\Sigma$ in matrix $W$ on $L$ \eqref{eq:Sigmacov}, both of which are in $\mathcal{H}_L$. Second, the magnification factor $\sigma$ multiplies most of the decision variables of the optimization. Thus neither $L$ nor $\sigma$ can be taken as variable in the optimization. We solve this by applying an iterative algorithm over both $L$ and $\sigma$ that leverages the structure presented in Figure \ref{fig:sigmashape}. If we select a large enough value for the magnification parameter $\sigma$, any choice of $L$ easily satisfies $\mathcal{H}_L$ by creating a very large ellipsoidal outer bound. We satisfy the other constraints of the optimization (the $\|H\|_2$ constraints) by selecting the initial value for $L$ as the $\|H\|_2$ optimal $L$ (from Theorem \ref{covbaseH}), hence satisfying $\mathcal{C}_h$ and $\mathcal{C}_L$. For these fixed values of $\sigma$, $L$, and $\Sigma$, the optimization \eqref{eq:semifinalconvex} is solved. The solution then provides a new value of $L$ - solved from the Ricatti equation in \eqref{eq:genricatti} - that minimizes the ellipsoidal bound while satisfying all constraints. Using the same value for $\sigma$ but the updated $L$ (and hence updated $\Sigma$), optimization \eqref{eq:semifinalconvex} is again solved, yielding another value for $L$. This iteration is repeated until $L$ has sufficiently converged. Once convergence is achieved, $\sigma$ is reduced and the process is repeated with the existing $L$ as the initial value for the $L$ iteration. At some point, the magnification factor $\sigma$ will be too small for any choice of $L$ to permit the ellipsoidal bound to contain the reachable set, hence the optimization \eqref{eq:semifinalconvex} will become infeasible. This is the stopping condition for the algorithm.  

Note that the convergence criteria for $L$ and the decrement amount for $\sigma$ are selected by the user. In principle, these should be small, but making them larger will allow the algorithm to require fewer iterations. The number of iterations for $L$ to converge tends to be quite small (typically 2-3). In practice, the decrement of $\sigma$ can also be accomplished through a bisection algorithm which looks for the smallest value of $\sigma$ that makes the optimization \eqref{eq:semifinalconvex} feasible.

Recall that, in general, solving the Ricatti equation \eqref{eq:genricatti} yields up to $2n \choose n$ solutions. We assert that there will always exist at least one real solution due to matrices $\mathbf{P}_L$, $\mathbf{P}$, $\mathcal{C}_L$, and $\mathcal{C}$ being positive definite. To chose between the real solutions, we choose the one that yields the smallest ellipsoidal outer bound, as determined by Lemma \ref{lem:analysis} (which provides a tighter approximation of the reachable set, since it does not require the additional linearization steps).

Finally, the stability of the closed loop system is implicitly guaranteed if $\mathcal{H}_L>0$ and $\mathcal{C}_L>0$ (see Appendix \ref{stab}).
\hspace*{\fill}~$\blacksquare$
\begin{remark}\label{rmk:initialguess}
In practice, the quality of the solution is related to the initial guess for $L$. For the cases where $\bar{\gamma}$ is close to $\gamma^*$ we use $L$ from Theorem \ref{covbaseH} as the initial guess. Otherwise when $\bar{\gamma}$ is far from $\gamma^*$ we solve the problem in multiple steps increasing $\bar{\gamma}$ gradually from $\gamma^*$ at each step solving Theorem \ref{semifinalconvex} and using its solution as the initial values for the next step where $\bar{\gamma}$ is increased. 
\end{remark}
\section{Resolving the Nonlinearity}
The approach offered in Theorem \ref{semifinalconvex} relies on an iterative scheme to avoid the nonlinearities surrounding the observer gain matrix $L$. In each iteration, the solution of the convex optimization in \eqref{eq:semifinalconvex} is used to find $L$ using the Ricatti equation in \eqref{eq:genricatti} and the corresponding residual covariance $\Sigma$ using the Lyapunov equation in \eqref{eq:estcov}. To eliminate the iterative approach, both of these nonlinearities must be linearized and absorbed into the convex optimization. A final nonlinearity is the existance of the  $-\mathcal{P}_3L$ term within $\mathcal{H}_L$ - specifically $B_L$ in \eqref{eq:newboundmatrices}. We accomplish this linearization through the careful selection of additional structure, which we will show only marginally reduces the quality of the solutions found.

\begin{structure} \label{structure:e}
Here $\mathcal{Q}_e=\mathcal{P}_3^{-1}$ is the shape matrix of the estimation error bounding ellipsoid. Again there are an infinite number of choices for its orientation and, following the same approach taken in Imposed Structure \ref{structure:xxhat}, here we choose the orientation that matches that of the estimation error covariance in the absence of attack $\mathcal{P}_3=\frac{1}{\sigma}\mathbf{P}_e^{-1}$.
\end{structure}

\begin{structure} \label{Pxxhat}
The final linearization assumption is to search for solutions on the manifold that satisfies $\mathbf{P}_{\hat{x}}=\mathbf{P}_{x\hat{x}}$. This is the most restrictive of the four structures that have been imposed to linearize the problem, however, we will see it does not greatly impact the conservatism of the optimization. One of the direct effects of this structure is that the generalized algebraic Ricatti equation \eqref{eq:genricatti}, one of the key nonlinearities, is directly linearized. However, it is this final structure that ultimately removes all three nonlinearities present in Theorem \ref{semifinalconvex}.
\end{structure}

From Imposed Structure \ref{Pxxhat} and the identity $\mathbf{P}\mathbf{Q}=I$ - see \eqref{eq:PinvQ} - we can show that $\mathbf{P}_e^{-1}=\mathbf{Q}_1=-\mathbf{Q}_{12}$ (see Appendix \ref{derivation}). The relationship between $\mathbf{Q}_{1}$ and $\mathbf{Q}_{12}$ is primarily what linearizes the Ricatti equation in \eqref{eq:genricatti}. The relationship between $\mathbf{P}_e$ and $\mathbf{Q}_1$, combined with Imposed structure \ref{structure:e}, enables us to replace the $-\mathcal{P}_3L$ nonlinearity with $\frac{1}{\sigma}X$. Therefore we can re-write matrices $A_L, B_L, \mathcal{P}_L$ as,
\begin{equation}
\begin{aligned}
&A_L=\begin{bmatrix}\sigma (F\mathbf{P}_x+GY) & F & 0\\ Z & \frac{1}{\sigma}(\mathbf{Q}_1F+XC) & -\frac{1}{\sigma}XC\\ 0 & 0 & \frac{1}{\sigma}\mathbf{Q}_1F\end{bmatrix},\\
&B_L=\begin{bmatrix*}[r]I\ \ &0\ \ \\\frac{1}{\sigma}\mathbf{Q}_1&\frac{1}{\sigma}X\\ \frac{1}{\sigma}\mathbf{Q}_1& \frac{1}{\sigma}X\end{bmatrix*}, \quad \mathcal{P}_L=\begin{bmatrix}\sigma \mathbf{P}_x &I&0\\I&\frac{1}{\sigma}\mathbf{Q}_1&0\\0&0&\frac{1}{\sigma}\mathbf{Q}_1\end{bmatrix}.
\end{aligned}
\end{equation}


Finally, the relationship between $\mathbf{P}_e$ and $\mathbf{Q}_1$ also help to integrate the Lyapunov equation in \eqref{eq:estcov} into the convex optimization. Lemma \ref{SigmaXQ12} relates the matrix $\Pi=\Sigma^{-1}/\alpha$ with the existing decision variables. This requires us, like the system and sensor noises, to truncate the estimation error distribution at some confidence level, measured by $\bar{e}$.
\begin{lemma}\label{SigmaXQ12}
Consider the truncated Gaussian system noise and measurement noise with $\nu_k^TR_1^{-1}\nu_k\leq \bar{\nu}$ and $\eta_k^TR_2^{-1}\eta_k\leq \bar{\eta}$ and their corresponding Gaussian estimation error and residual with $e_k^T\mathbf{P}_e^{-1}e_k\leq \bar{e}$ and $r_k^T \Pi r_k\leq 1$. The positive definite matrix $\Pi$, can be expressed as $\Pi=\frac{\Sigma^{-1}}{\alpha}$ where $\Sigma$ comes from \eqref{eq:Sigmacov} and \eqref{eq:estcov}, if,
\begin{align}
    &\mathcal{X}_L=\begin{bmatrix} \mathbf{Q}_{1}-(\bar{e}+\bar{\eta})C^T \Pi C & -(\bar{e}+\bar{\eta})C^T \Pi \\ -(\bar{e}+\bar{\eta})\Pi C & R_2^{-1}-(\bar{e}+\bar{\eta})\Pi\end{bmatrix}\geq 0. \label{eq:upboundpi}\\
    & \mathcal{S}_L=\resizebox{0.75\hsize}{!}{$\begin{bmatrix}
    \mathbf{Q}_{1} & \mathbf{Q}_{1}F+XC & \mathbf{Q}_{1}R_1 &XR_2\\ (\mathbf{Q}_{1}F+XC)^T & \mathbf{Q}_{1} & 0 & 0 \\
    R_1\mathbf{Q}_{1} &0& R_1 & 0 \\ 
    R_2X^T & 0 & 0 &  R_2
    \end{bmatrix}$} \geq 0. \label{eq:SL}
\end{align}
where $\bar{e}$ is computed from the lower incomplete Gamma function such that $\text{Pr}[e_k^T\mathbf{P}_e^{-1}e_k \leq \bar{e}] = p_e$, and $p_e$ is the desired trucation probability.
\end{lemma}
\textit{Proof:}
The inequality $\mathcal{H}_L$ already provides a lower bound on $\Pi$. We now add an additional upper constraint to sandwich and fully constrain $\Pi$. We exploit the fact that $r_k^T \Pi r_k\leq1$, and $r_k=Ce_k+\eta_k$, therefore we conclude,
\begin{equation}
\Phi_1 = \begin{bmatrix}e_k^T & \eta_k^T \end{bmatrix} \begin{bmatrix}C^T\Pi C & C^T \Pi \\ \Pi C & \Pi\end{bmatrix}\begin{bmatrix}e_k\\ \eta_k \end{bmatrix} \leq 1.
\end{equation}
In addition, for the estimation error in the absence of attack we have, $e_k^T \mathbf{P}_e^{-1} e_k \leq \bar{e}$, where $\bar{e}$ is chosen to contain most of the Gaussian estimation error distribution (this quadratic form follows a chi-squared distribution, therefore, $\bar{e}$ can be chosen by the lower incomplete Gamma function). For the measurement noise we have, $\eta_k^T R_2^{-1} \eta_k \leq \bar{\eta}$, therefore, together we have
\begin{equation}
\Phi_2 = \begin{bmatrix}e_k^T & \eta_k^T \end{bmatrix} \begin{bmatrix}\mathbf{P}_e^{-1} & 0 \\ 0 & R_2^{-1}\end{bmatrix}\begin{bmatrix}e_k\\ \eta_k \end{bmatrix} \leq \bar{e}+\bar{\eta}.
\end{equation}
Hence, we can provide a upper bound for $\Pi$ since $\Phi_1\leq 1$, then $(\bar{e}+\bar{\eta})\Phi_1 \leq \Phi_2$. Thus,
\begin{equation}
\begin{bmatrix}\mathbf{P}_e^{-1}-(\bar{e}+\bar{\eta})C^T \Pi C & -(\bar{e}+\bar{\eta})C^T \Pi \\ -(\bar{e}+\bar{\eta})\Pi C & R_2^{-1}-(\bar{e}+\bar{\eta})\Pi\end{bmatrix}\geq 0
\end{equation}
which can be written as \eqref{eq:upboundpi}. We now have a upper bound on $\Pi$ in terms of $\mathbf{Q}_1$, however, $\mathbf{Q}_1$ is not itself constrained. To do this, we relax the Lyapunov equation \eqref{eq:estcov} equality to inequality,
\begin{equation}\label{eq:estcovrelax}
\mathbf{P}_e-(F-LC)\mathbf{P}_e(F-LC)^T-R_1-LR_2L^T \geq0.
\end{equation}
We apply the Schur Complement to receive
\begin{equation}
\mathcal{S}=\begin{bmatrix}\mathbf{P}_e-LR_2L^T-R_1 & (F-LC)\mathbf{P}_e\\ \mathbf{P}_e(F-LC)^T & \mathbf{P}_e\end{bmatrix}\geq 0.
\end{equation}
Using the transformation $T_4=\text{diag}\begin{bmatrix}\mathbf{P}_e^{-1} & \mathbf{P}_e^{-1}\end{bmatrix}$ gives
\begin{equation}
    \mathcal{S}_L=T_4 \mathcal{S} T_4^T=\begin{bmatrix}\mathbf{P}_{eL}-R_{eL} & A_{eL}\\ A_{eL}^T & \mathbf{P}_{eL}\end{bmatrix}\geq 0,
\end{equation}
where,
\begin{align}
    & \mathbf{P}_{eL}=\mathbf{P}_e^{-1}=\mathbf{Q}_1,\\
    & R_{eL}=\mathbf{P}_e^{-1}(R_1+LR_2L^T)\mathbf{P}_e^{-1}= \mathbf{Q}_{1}R_1\mathbf{Q}_{1}+XR_2X^T,\nonumber\\
    &A_{eL}=\mathbf{P}_e^{-1}(F-LC)=\mathbf{Q}_{1}F+XC.\nonumber
\end{align}
Applying the Schur Complement again yields the expression in \eqref{eq:SL}.
\hspace*{\fill}~$\blacksquare$

With these combined linearization steps, we can provide Theorem \ref{finalconvex}.


\begin{theorem}\label{finalconvex}
Consider a LTI system \eqref{eq:dLTIsystem} with desired output covariance constrained $\|H\|_2$ gain $\bar{\gamma}$ \eqref{eq:originalineq}, chi-squared detector threshold $\alpha$ \eqref{chisquared} and zero-alarm stealthy attacker \eqref{eq:zeroalarmdef}. If there exist $a, a_2 \in [0,1)$ then the solution of the following convex optimization provides the optimal observer $L$ and controller $K$ gain matrices that minimize the set of states reachable by an attacker while maintaining an OCC $\|H\|_2$ gain no bigger than $\bar{\gamma}$,
\begin{equation}\label{eq:finalconvex}
\begin{aligned}
&\left\{ \begin{aligned}
&\min_{{a_1,\mathbf{P}_x,\mathbf{Q}_1,\Pi}\atop{X,Y,Z}} \mathbf{tr}(\sigma \mathbf{P}_x)\\
&\qquad\text{s.t.}\quad 0\leq a_1< 1, \quad a_1+a_2 \geq a,\\
&\qquad\qquad\mathcal{H}_L\geq0,\  \mathcal{C}_{h} \leq 0,\ \mathcal{C}_L \geq 0,\\
&\qquad\qquad \mathcal{S}_L\geq0,\ \mathcal{X}_L\geq0.
\end{aligned}\right.
\end{aligned}
\end{equation}
This optimization is solved for a specific value of $\sigma$, which is optimized through a bisection algorithm.
\end{theorem}

\section{Case study} 
We consider a LTI system (with matrices given below) for this study with the chi-squared detector tuned to a false alarm rate $\mathcal{A}=0.05$ (5\%), system noise truncated with $\bar{p}=95\%$, and a worst acceptable OCC $\|H\|_2$ gain of $\bar{\gamma}=8.75$. We use CVX to solve the convex optimizations \cite{cvx}.
\begin{equation*}
\begin{aligned}
&F= \begin{bmatrix*}[r]
    1.0444&   -0.1409\\
    0.3001&    0.6327
     \end{bmatrix*},\ 
G=\begin{bmatrix}
       2 & 3 \\
       1 & 1
     \end{bmatrix},\ 
C= \begin{bmatrix}
      2 & 2 \\
       1 & 2
     \end{bmatrix}, \\
&R_1= \begin{bmatrix*}[r]
       0.0183&   -0.0218\\
       -0.0218&    0.0261
     \end{bmatrix*}, \ 
R_2=\begin{bmatrix}
       0.0018& 0.0031\\
        0.0031& 0.0096
     \end{bmatrix}
\end{aligned}
\end{equation*}
We start by computing the optimal OCC $\|H\|_2$ gain $\gamma^*$ using Theorem \ref{covbaseH},
\begin{equation}
\gamma^*=\sqrt{\frac{ \mathrm{tr} \big( C \  \mathbf{P}_x \ C^T\big) + \mathrm{tr}(R_2) }{\mathrm{tr}(R_2)+\mathrm{tr}(R_1)}}=1.5673.
\end{equation}
The solution for gain matrices $L$ and $K$ based on the generalized algebraic Ricatti equation \eqref{eq:genricatti}, returns ${4\choose 2}=6$ different answers of which two are real valued. Of the real solutions,
\begin{equation}\label{eq:covbaseHgains}
L=\begin{bmatrix*}[r]1.0085 & -0.9780 \\ -0.0139 & 0.2664\end{bmatrix*}, \quad  K=\begin{bmatrix*}[r]0.1273 &  -2.0544 \\ -0.4303 &  1.4190\end{bmatrix*},
\end{equation}
provides the smallest reachable set according to Lemma \ref{lem:analysis}. To examine the accuracy of Theorem \ref{covbaseH}, we compare the state covariance that is the solution of the Lyapunov equation \eqref{eq:xcovariance} with the designed gains in \eqref{eq:covbaseHgains}, with the decision variable $\mathbf{P}_x$ of the optimization. In this case, the (entry-wise) mean absolute error (MAE) between these matrices is less than $0.00001$, justifying that the relaxed inequality in Theorem \ref{covbaseH} recovers an exact Lyapunov equation solution.

The open loop OCC $\|H\|_2$ gain is found using Lemma \ref{lem:openloopgamma} to be,
\begin{equation}
\gamma_0=\sqrt{\frac{ \mathrm{tr} \big( C^TC \  \mathbf{P}_x \ \big) + \mathrm{tr}(R_2) }{\mathrm{tr}(R_2)+\mathrm{tr}(R_1)}}=10.1899.
\end{equation}

Next we use the iterative approach from Theorem \ref{semifinalconvex} with convergence criteria $\epsilon=0.03$ for the gain matrix $L$. We use bisection to find the smallest feasible $\sigma$ (stopping tolerance of 0.01), starting with $\sigma_{min}=0.1$ and $\sigma_{max}=10^{6}$. In addition we increase $\bar{\gamma}$ gradually from $\gamma^*=1.57$ until $\bar{\gamma}=8.75$ by steps of $0.1$ (see Remark \ref{rmk:initialguess}). The optimal gains are,
\begin{equation}
    L=\begin{bmatrix} 0.0956  & -0.1248\\-0.1010 & 0.1321 \end{bmatrix}, \quad  K=\begin{bmatrix} 0.1440 & -2.0390\\ -0.4441  &  1.4063 \end{bmatrix},
\end{equation}
with corresponding magnification factor $\sigma=400.91$.
Given these gains the actual performance is computed as $\gamma=6.91$, which is computed by substituting the optimal gains into and then solving the Lyapunov equation \eqref{eq:xcovariance} and then computing $\gamma$ from \eqref{eq:nonconvexgamma}. This discrepancy between the input $\bar{\gamma}$ and the outcome $\gamma$ is the price of the linearization, caused by the competing LMI constraints (see Appendix \ref{gamadif}).

Finally we optimize the gains $L$ and $K$ using the fully convexified method in Theorem \ref{finalconvex}, which results in,
\begin{equation}\label{eq:fullyconvexgains}
    L=\begin{bmatrix} 0.1274  & -0.1737\\-0.2019 & 0.2872 \end{bmatrix}, \quad  K=\begin{bmatrix} 0.1902  & -1.9945\\ -0.4757 & 1.3759 \end{bmatrix}.
\end{equation}
The actual performance is $\gamma=6.81$ (as above). The bisection is configured as above and returns $\sigma=2471.42$ as the smallest feasible value of the magnification factor. 
To examine the accuracy of these linearization method we compare the state covariance of the system which is the solution of the Lyapunov equation (as discussed above) with the designed gains in \eqref{eq:fullyconvexgains}, with the decision variable $\mathbf{P}_x$ of the optimization. The mean absolute error between these two matrices is $0.04$. As before, this discrepancy is the result of the linearization (see Appendix \ref{inf})

Figure \ref{fig:elips875} compares the ellipsoidal bounds provided by the solutions above including Theorem \ref{finalconvex} (red) and Theorem \ref{semifinalconvex} (blue) for the worst allowable performance $\bar{\gamma}=8.75$, as well as Theorem \ref{covbaseH} (black) for optimal performance $\gamma=\gamma^*$. In this figure the ellipsoidal bounds are from Lemma \ref{lem:analysis} with optimal gains provided from Theorem \ref{finalconvex}, Theorem \ref{semifinalconvex}, and Theorem \ref{covbaseH}. This figure demonstrates that despite the similarities between Robust Contol and Resilient Control design, the $\|H\|_2$ optimal solution is not the optimal security solution. It also shows that the fully convexified solution evidences only marginally degraded solution quality, despite the linearization steps.

\begin{figure}[t]
    \begin{center}
    \includegraphics[width=1\linewidth]{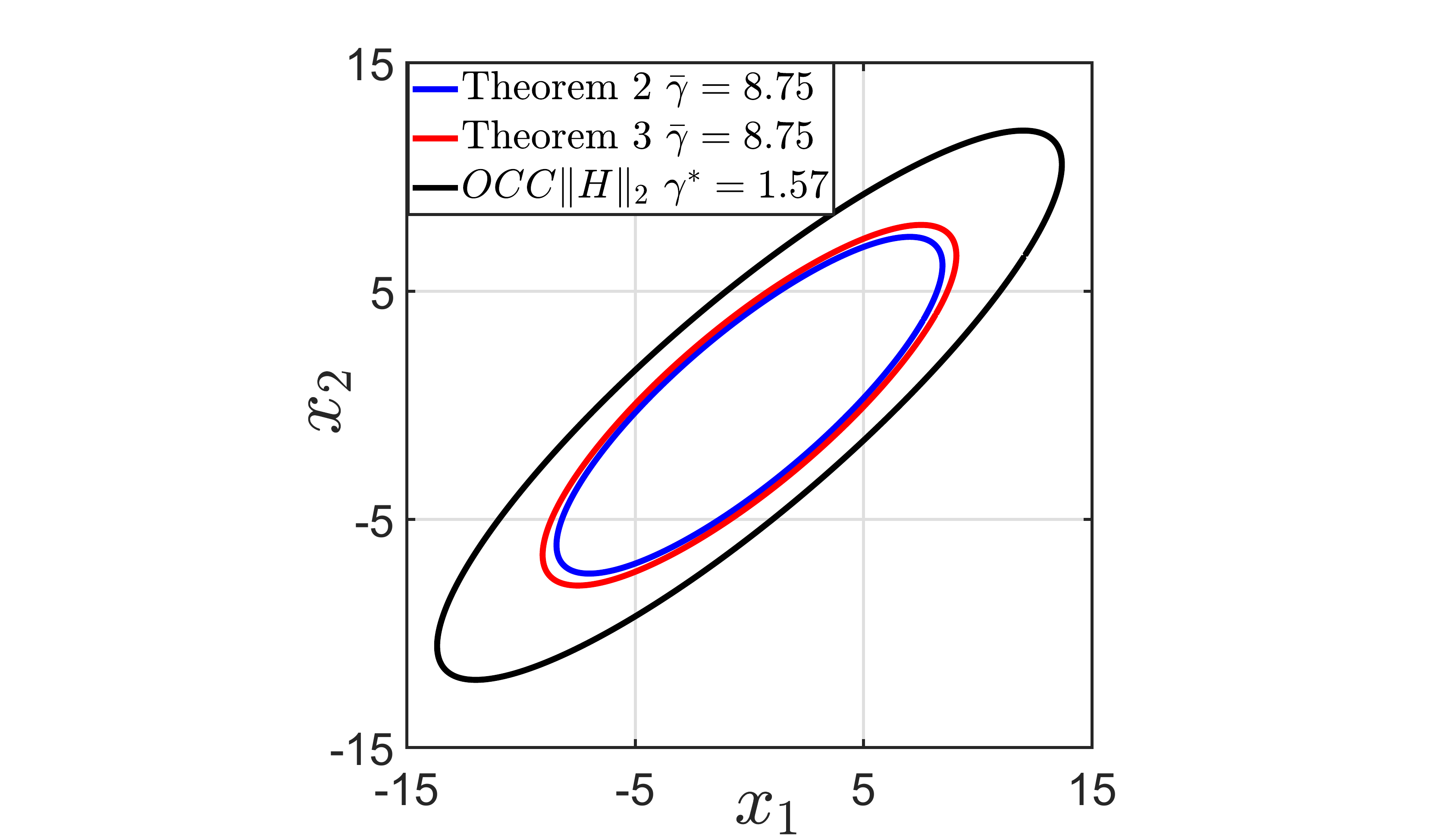}
    \end{center}
    \caption{A comparison between the ellipsoidal bounds of the optimized reachable sets from Theorems \ref{semifinalconvex} and \ref{finalconvex} for desired performance $\bar{\gamma}=8.75$ along with that of the OCC $\|H\|_2$ optimal solution.}
    \label{fig:elips875}
\end{figure}

\begin{figure}[t]
    \begin{center}
    \includegraphics[width=1\linewidth]{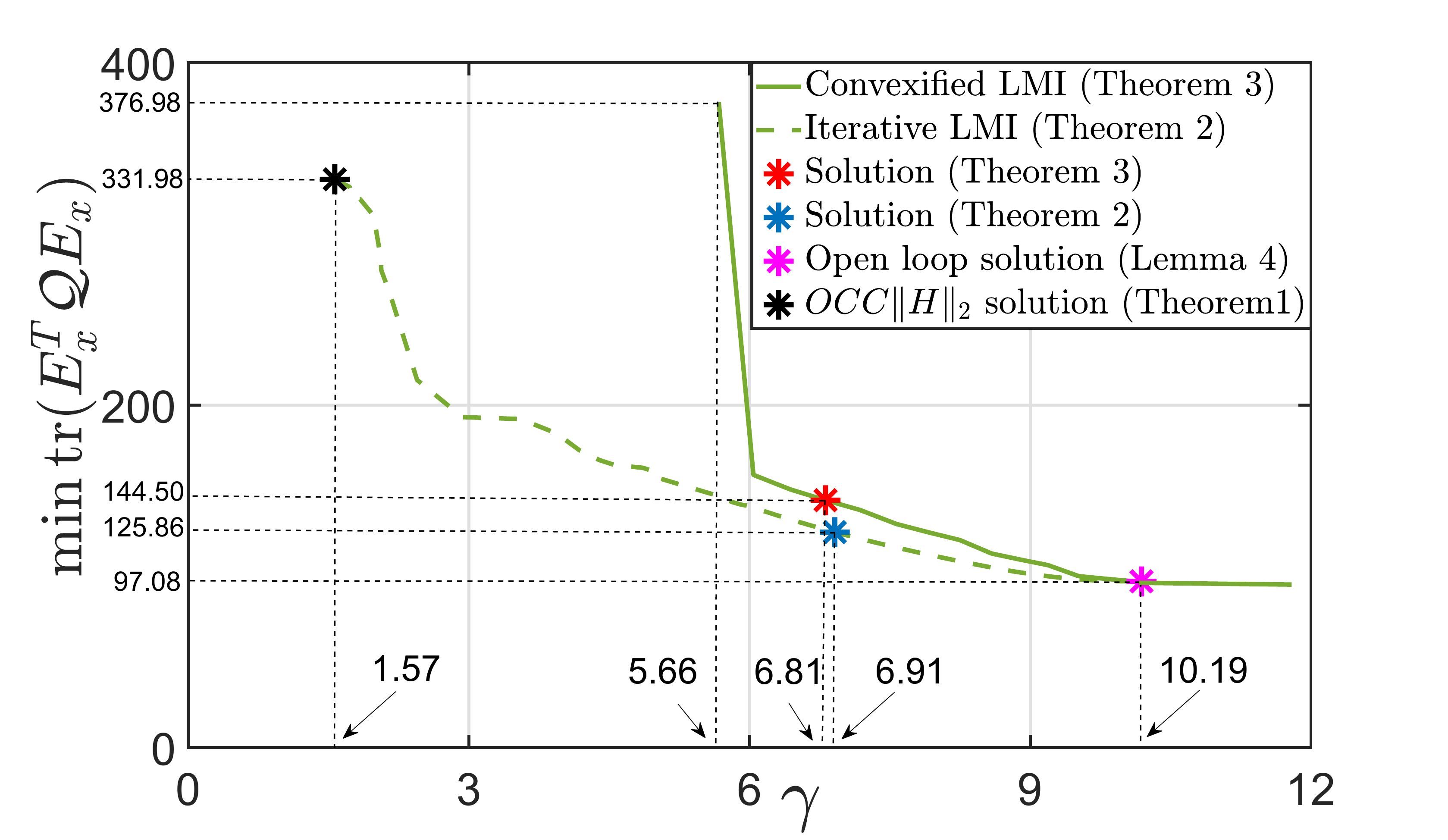}
    \end{center}
    \caption{Performance-Security curve for the proposed iterative semi-definite optimization approach (dashed green) compared with fully convexified version (green). The extra steps taken to linearize the problem make the fully convex optimization feasible only for a portion of the entire trade off interval.}
    \label{fig:perfsec}
\end{figure}

In Figure \ref{fig:perfsec} we compute the optimal gains for a wide range of worst allowable OCC $\|H\|_2$ gains using Theorem \ref{semifinalconvex} and Theorem \ref{finalconvex}. Given the optimal gains, $L$ and $K$ for each, we plot the objective function of Lemma \ref{lem:analysis} with respect to the actual performance, $\gamma$. We use the objective function of Lemma \ref{lem:analysis} as a proxy for security where, smaller values imply better security. This plot demonstrates that there is, indeed, a trade off between  security and performance. The nonlinearity and steep slope near $\gamma^*$ indicates that dramatic improvements in security can be gained by marginal concessions in performance. The relationship between these two properties, captured by this plot, should be used in resilient control system design. 

Figure \ref{fig:perfsec} shows that the solution of Theorem \ref{finalconvex} does not cover the entire trade off interval and the problem is infeasible on the interval $\gamma \in [\gamma^*,\ \gamma_c]$, where $\gamma_c=5.66$ for this scenario. The manifold $\mathbf{P}_{\hat{x}}=\mathbf{P}_{x\hat{x}}$ imposed by Imposed Structure \ref{structure:xxhat} is not able to provide a feasible solution for entire trade off interval, satisfying both the stability constraint in both attacked system LMI ($\mathcal{H}_L$) and attack free LMI ($\mathcal{C}_L$). There exists an infimum for the optimal threshold on this manifold, which is computed in Appendix \ref{inf}. 

\section{Conclusion}
This paper presents a set of tools to design the feedback controller and observer gains for observer-based feedback control with the aim to minimize the effect sensor falsification attacks. As past work has observed, there is a necessary trade-off between ensuring performance and minimizing the effect of the attacker. Here the attacker impact is quantified as the set of states reachable through the action of the attacker and the nominal closed-loop performance is specified by an output covariance constrained $\|H\|_2$ gain. We frame this problem as an LMI wrapped in an iterative algorithm as well as a fully convexified optimization and a large part of the effort here is to linearize the constraints involved. Along the way, we also contribute a convex optimization to design optimal OCC $\|H\|_2$ gains.


\bibliographystyle{plain}        
\bibliography{main}           



\appendix
\section{Validity of Relaxing the Lyapunov Equation}\label{apdx:lyapunov_relax}
Here we show more precisely the objective function $\mathbf{tr}(C\mathbf{P}_xC^T)$ forces the relaxed inequality constraint \eqref{eq:replacement} to its boundary (equality), hence the optimization finds a solution that satisfies the original Lyapunov equation \eqref{eq:xcovariance}. To see this, consider the function, 
\begin{equation}
    f(\mathbf{P})=\mathbf{P}-\hat{A}\mathbf{P}\hat{A}^T-\hat{R}.
\end{equation}
and assume there exists matrices $\mathbf{P}\geq 0$ a solution to the optimization \eqref{eq:gammaopt} and $\bar{\mathbf{P}}\geq 0$ a solution of the Lyapunov equation \eqref{eq:xcovariance} such that,
\begin{align}
f(\mathbf{P})&=\mathbf{P}-\hat{A}\mathbf{P}\hat{A}^T-\hat{R} \geq 0,\\
f(\bar{\mathbf{P}})&=\bar{\mathbf{P}}-\hat{A}\bar{\mathbf{P}}\hat{A}^T-\hat{R} = 0.
\end{align}
The difference $f(\mathbf{P})-f(\bar{\mathbf{P}})\geq 0$ is positive semi-definite, so it also satisfies a Lyapunov equation with stable $\hat{A}$ which means,
\begin{equation}
(\mathbf{P} - \bar{\mathbf{P}})-\hat{A}(\mathbf{P}-\bar{\mathbf{P}})\hat{A}^T - \bar{R} = 0,
\end{equation}
for some $\bar{R}\geq 0$. As the solution of a Lyapunov equation the difference $\mathbf{P}-\bar{\mathbf{P}}\geq 0$ is positive semi-definite, and so the first block corresponding to the states $\mathbf{P}_x - \bar{\mathbf{P}}_x \geq 0$ is positive semi-definite. Thus,
\begin{align}
C\mathbf{P}_x C^T &\geq C\bar{\mathbf{P}}_xC^T, \\
\mathbf{tr}(C\mathbf{P}_xC^T) &\geq \mathbf{tr}(C\bar{\mathbf{P}}_xC^T). \label{eq:alwayssmaller}
\end{align}
In the absence of no constraints other than $\mathcal{C}_L$, this inequality demonstrates that if the objective function $\mathbf{tr}(C\mathbf{P}_xC^T)$ is minimized, that the solution satisfies the Lyapunov equation \eqref{eq:xcovariance}, i.e., that the optimization forces the inequality in \eqref{eq:replacement} to equality. 

\section{Proof of Stability}\label{stab}   
We show that the LMI constraints $\mathcal{H}_L>0$ and $\mathcal{C}_L>0$ imply that the attacked system and nominal system are stable.

For $\mathcal{H}_L$, $\mathcal{H}_L>0$ implies $\mathcal{H}>0$ which implies $\mathcal{P}>0$ and $a\mathcal{P}-A^T\mathcal{P}A>0$. Since $a\in [0,1)$, $(1-a)\mathcal{P}+a\mathcal{P}-A^T\mathcal{P}A = \mathcal{P}-A^T\mathcal{P}A>0$. Thus, $A^T$, and hence, $A$ is stable.

Similarly, for $\mathcal{C}_L$, $\mathcal{C}_L>0$ implies $\mathcal{C}>0$ which implies $\mathbf{P}>0$ and $\mathbf{P}-\hat{A}\mathbf{P} \hat{A}^T-\hat{R}>0$. Therefore, $\mathbf{P}-\hat{A}\mathbf{P} \hat{A}^T >0$, which makes $\hat{A}$ stable as well.

\section{Manifold of $\mathbf{P}_{\hat{x}}=\mathbf{P}_{x\hat{x}}$}\label{derivation} 
We know that $\mathbf{P}\mathbf{Q}=I$ which means,
\begin{align}
&\mathbf{P}_x\mathbf{Q}_1+\mathbf{P}_{x\hat{x}}\mathbf{Q}_{12}^T=I\label{n1}\\
&\mathbf{P}_x\mathbf{Q}_{12}+\mathbf{P}_{x\hat{x}}\mathbf{Q}_2=0\label{n2}\\
&\mathbf{P}_{x\hat{x}}^T\mathbf{Q}_1+\mathbf{P}_{\hat{x}}\mathbf{Q}_{12}^T=0\label{n3}\\
&\mathbf{P}_{x\hat{x}}^T\mathbf{Q}_{12}+\mathbf{P}_{\hat{x}}\mathbf{Q}_2=I.\label{n4}
\end{align}
if we assume the points located on manifold $(\mathbf{P}_{\hat{x}}=\mathbf{P}_{x\hat{x}})$, Based on \eqref{n3} we can conclude, $\mathbf{Q}_1=-\mathbf{Q}_{12}$ and based on \eqref{n1} we have
\begin{equation}\label{1st}
\mathbf{P}_{\hat{x}} = \mathbf{P}_x-\mathbf{Q}_1^{-1}.
\end{equation}
Based on $e_k=x_k-\hat{x}_k$,
\begin{equation}
    \mathbf{P}_e=\mathbf{P}_x+\mathbf{P}_{\hat{x}}-\mathbf{P}_{x\hat{x}}-\mathbf{P}_{x\hat{x}}^T,
\end{equation}
where on the mentioned manifold,
\begin{equation}\label{2nd}
    \mathbf{P}_e=\mathbf{P}_x-\mathbf{P}_{\hat{x}}.
\end{equation}
Comparing \eqref{1st} and \eqref{2nd} we can conclude, $\mathbf{P}_e^{-1}=\mathbf{Q}_1$.

\section{Infimum value of desired performance on the manifold $\textbf{P}_{\hat{x}}=\textbf{P}_{x\hat{x}}$ }\label{inf}
To compute this infimum value, it suffices to introduce $\tau=\bar{\gamma}^2$, and obtain an optimal solution of $\tau$ from the convex optimization,
\begin{equation}\label{eq:gammac}
\begin{aligned}
&\left\{ \begin{aligned}
&\min_{{a_1,\mathbf{P}_x,\mathbf{Q}_1,\Pi}\atop{X,Y,Z, \tau}} \tau \\
&\qquad\text{s.t.}\quad 0\leq a_1< 1, \quad a_1+a_2 \geq a,\\
&\qquad\qquad\mathcal{H}_L\geq0,\  \mathcal{C}_{h} \leq 0,\ \mathcal{C}_L \geq 0,\\
&\qquad\qquad \mathcal{S}_L\geq0,\ \mathcal{X}_L\geq0,
\end{aligned}\right.
\end{aligned}
\end{equation}
where $a,a_2 \in [0,\ 1]$ are introduced as a grid search variables. Here we have replaced the objective function that minimizes the reachable set with an objective that minimizes $\bar{\gamma}$ without penalty for increasing the reachable set size (either through the direct decision variables or through $\sigma$). The scaling factor $\sigma$ for this optimization is fixed as a large value (to make the optimization feasible, but there is no need to solve for the smallest value of $\sigma$ here).
For this case, the optimal $\bar{\gamma}=\bar{\gamma}_c=7.75$ and then based on the optimal gains $L$ and $K$ the actual performance is $\gamma_c=5.66$. The jump at $\gamma_c$ in Figure \ref{fig:perfsec} is because the system becomes marginally stability.

\section{Evaluation of optimal performance from Theorem \ref{semifinalconvex} and \ref{finalconvex} }\label{gamadif}

The cost we pay for linearization in this work is that the optimizations in Theorems \ref{semifinalconvex} and \ref{finalconvex}, unlike Theorem \ref{covbaseH}, cannot guarantee that constraints $\mathcal{C}_h$ and $\mathcal{C}_L$ meet their boundary perfectly due to the addition of other competing inequalities. Therefore, the solution of the decision variable $\mathbf{P}_x$ is slightly different from what would be computed from the Lyapunov equation \eqref{eq:xcovariance} using the solution gains $K$ and $L$. Consequently the performance $\gamma$ is different from $\bar{\gamma}$. 
The $\mathcal{C}_h$ constraint, however, ensures that the actual performance, $\gamma$, will be smaller than the specified performance, $\bar{\gamma}$.

\begin{figure}[t]
    \centering
    \includegraphics[width=1\linewidth]{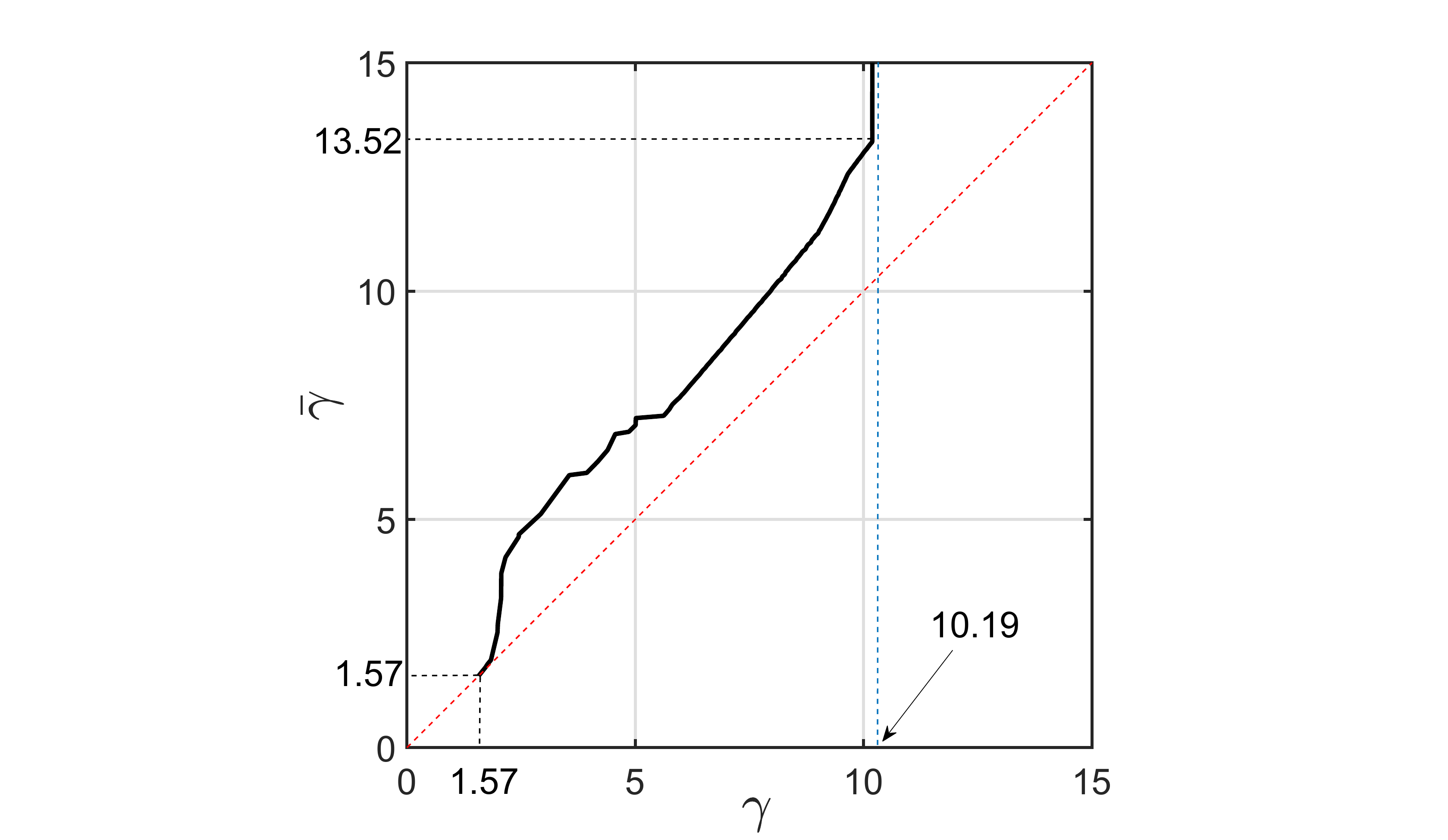}
    \caption{Optimal performance versus specified performance in the solutions of Theorem \ref{semifinalconvex}. In the vicinity of $\gamma=\gamma_0$, we see that $\bar{\gamma}$ diverges.}
    \label{fig:theorem2}
\end{figure}

\begin{figure}[t]
    \centering
    \includegraphics[width=1\linewidth]{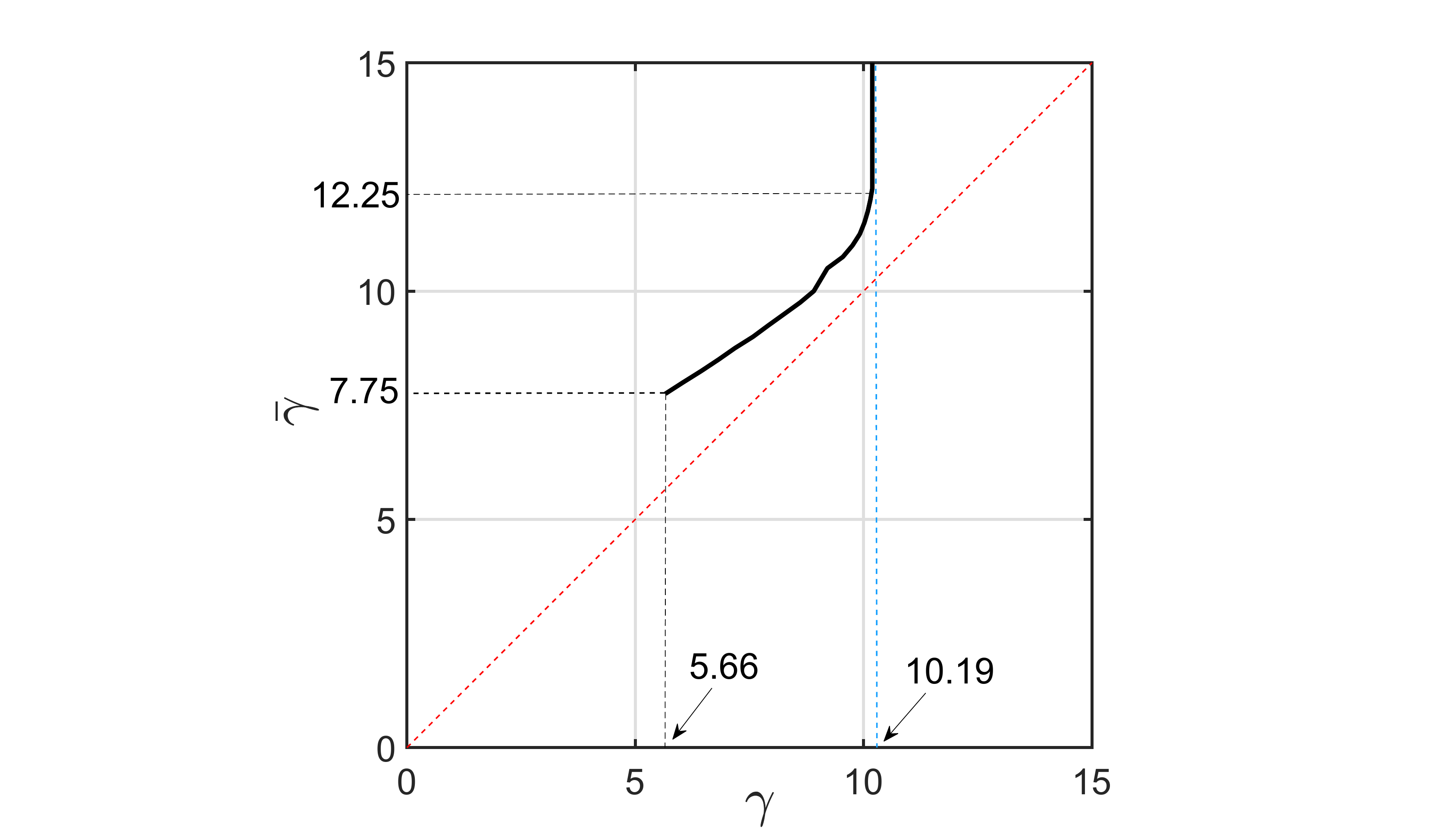}
    \caption{Optimal performance versus specified performance in the solutions of Theorem \ref{finalconvex}. Here $\bar{\gamma}_c=7.75$ and $\gamma_c=5.66$. In the vicinity of $\gamma=\gamma_0$, we see that $\bar{\gamma}$ diverges.}
    \label{fig:theorem3}
\end{figure}

Figure \ref{fig:theorem2} and Figure \ref{fig:theorem3} depict the difference between $\gamma$ and $\bar{\gamma}$ in the iterative and fully convexified optimizations, respectively. In addition, these figures can be used as a tool to determine what desired performance should be imposed on the linearized problem in order to receive expected performance in original nonlinear problem. 



\end{document}